\documentclass{ws-ijmpb}

\newcommand{\rvec}{{\bf r}} 
\newcommand{\vvec}{{\bf v}} 
\newcommand{\bvec}{{\bf b}}
\newcommand{\kvec}{{\bf k}}
\newcommand{\qvec}{{\bf q}}
\newcommand{\Rvec}{{\bf R}}

\newcommand{\Avec}{{\bf A}}
\newcommand{\Bvec}{{\bf B}}
\newcommand{\Mvec}{{\bf M}}
\newcommand{\Evec}{{\bf E}}
\newcommand{\mvec}{{\bf m}}
\newcommand{\home}{{\bf 0}}
\newcommand{\Jvec}{{\bf J}}
\newcommand{\Pvec}{{\bf P}}

\newcommand{\ket}[1]{|#1\rangle}
\newcommand{\bra}[1]{\langle#1|}
\newcommand{\me}[3]{\langle#1|#2|#3\rangle}
\newcommand{\orb}{_\text{orb}}
\newcommand{\park}{\partial_{\bf k}}
\newcommand{\mr}[2]{\multirow{#1}{*}{#2}}
\newcommand{\dr}{d^3\mspace{-2mu}r}
\newcommand{\dk}{d^3\mspace{-2mu}k}

\usepackage[superscript]{cite}
\usepackage{multirow}
\usepackage{hyperref}

\def\trimmarks{}

\begin{document}

\markboth{T. Thonhauser}{Theory of Orbital Magnetization in Solids}

%%%%%%%%%%%%%%%%%%%%%% Publisher's Area please ignore %%%%%%%%%%%%%%%%%%
%
\catchline{25}{00}{2011}{}{}
%
%%%%%%%%%%%%%%%%%%%%%%%%%%%%%%%%%%%%%%%%%%%%%%%%%%%%%%%%%%%%%%%%%%%%%%%%

\title{THEORY OF ORBITAL MAGNETIZATION IN SOLIDS}

\author{T. THONHAUSER}

\address{Department of Physics, Wake Forest University,\\
Winston-Salem, North Carolina 27109, USA.\\
http:/$\!$/www.wfu.edu/{\tt\~{}}thonhat\\
\email{thonhauser@wfu.edu}}

\maketitle

\begin{history}
\received{15 April 2011}
%\accepted{(Day Month Year)}
%\comby{(xxxxxxxxxx)}
\end{history}

\begin{abstract}
In this review article, we survey the relatively new theory of orbital
magnetization in solids---often referred to as the ``modern theory of
orbital magnetization''---and its applications. Surprisingly, while the
calculation of the orbital magnetization in finite systems such as atoms
and molecules is straight forward, in extended systems or solids it has
long eluded calculations owing to the fact that the position operator is
ill-defined in such a context. Approaches that overcome this problem
were first developed in 2005 and in the first part of this review we
present the main ideas reaching from a Wannier function approach to
semi-classical and finite-temperature formalisms. In the second part, we
describe practical aspects of calculating the orbital magnetization,
such as taking $k$-space derivatives, a formalism for pseudopotentials,
a single $k$-point derivation, a Wannier interpolation scheme, and DFT
specific aspects. We then show results of recent calculations on Fe, Co,
and Ni. In the last part of this review, we focus on direct applications
of the orbital magnetization. In particular, we will review how
properties such as the nuclear magnetic resonance shielding tensor and
the electron paramagnetic resonance $g$-tensor can elegantly be
calculated in terms of a derivative of the orbital magnetization.
\end{abstract}

\keywords{Magnetism; orbital magnetic moment and magnetization; solids;
periodic and extended systems; Berry phase; Wannier functions; Bloch
functions; semi-classical; finite temperature; pseudopotentials; Chern
invariant; nuclear magnetic resonance; electron paramagnetic resonance;
first-principles; density functional theory.}

%%%%%%%%%%%%%%%%%%%%%%%%%%%%%%%%%%%%%%%%%%%%%%%%%%%%%%%%%%%%%%%%%%%%%%%%
\section{Introduction}
\label{sec:introduction}
%%%%%%%%%%%%%%%%%%%%%%%%%%%%%%%%%%%%%%%%%%%%%%%%%%%%%%%%%%%%%%%%%%%%%%%%

Magnetism is one of the oldest phenomena known to mankind. In the
forms of lodestone and magnetite it is first mentioned as early as
4000~BC in Chinese writings.\cite{history_1} The name is often related
to the Greek province of Magnesia where the mineral was first mined, or
to the shepherd Magnes, ``the nails of  whose shoes and the tip of whose
staff stuck fast in a magnetick field while  he pastured his
flocks.''\cite{history_1,history_2} However, it was not until the
19$^\text{th}$ century that scientists such as Oersted, Amp{\`e}re,
Gauss, Faraday, and Maxwell took much of the mysticism out of magnetism
and put its theory on a stronger, classical scientific foundation.
Magnetism's quantum-mechanical origin was then developed during the
first half of the 20$^\text{th}$ century. Today, many standard textbooks
on the subject exist, courses in \emph{Electricity and Magnetism} are
required for many undergraduate and graduate students, and magnetism is
omnipresent in today's life through a vast amount of technical
applications and gadgets that rely on it. As such, it is rather
surprising that, until just recently, \emph{no} complete fundamental
theory for magnetism in solids existed.

Nowadays, it is well understood that magnetism in atoms and molecules
originates from two distinct sources related to quantum-mechanics: the
\emph{spin magnetic moment} and the \emph{orbital magnetic moment} of
the electrons.\footnote{Here, we are neglecting the magnetic moment
originating from the nuclear spins, which are usually smaller by several
orders of magnitude.} Rigorous theories for both parts exist at several
levels, and thus, the calculation of magnetic moments of atoms and
molecules has become a standard procedure. The situation in solids is
similar; the quantity of interest here is the magnetization, i.e.\ the
magnetic moment per volume. In parallel to atoms and molecules, the
magnetization in solids has two distinct contributions: the \emph{spin
magnetization} and the \emph{orbital magnetization}. However, while the
first can be computed easily, the latter has eluded exact calculation
due to the lack of an appropriate theory. It was not until 2005 that the
corresponding theory was developed. Note that already for several
decades the two contributions can be determined separately through a
combination of magneto-mechanical measurements.\cite{Meyer_61} In the
present review article we focus on this relatively new \emph{modern
theory of orbital magnetization in solids}, collecting the fundamental
derivations, its applications, and first results. 

The focus of this review article is entirely on the orbital contribution
to magnetism in solids---as the title emphasizes. For the purpose of
this review article we may refer to solids also as infinite, periodic,
or extended systems, or simply as crystals. The calculation of the
orbital contributions in finite systems, i.e.\ atoms and molecules, is
well established and is not the subject of this review; it can readily
be found in standard textbooks.

This review article is organized in the following way: In
Section~\ref{sec:importance}, we explain the importance of the orbital
magnetization and its relation to several other properties  of current
interest. Section~\ref{sec:theory} contains the theoretical framework
for the theory of orbital magnetization. First, in
Section~\ref{sec:problem}, we focus on the difference between finite
systems and solids and explain the mathematical difficulties that arise
in the context of periodic systems. Then, we show several different
derivations of the main theory, from a Wannier function approach in
Sections~\ref{sec:wannier} and \ref{sec:multi-band} to a semi-classical
(Section~\ref{sec:semi-classical}) and  finite-temperature
(Section~\ref{sec:finite-temperature}) formalism, which can be extended
to interacting systems. Practical aspects of calculating the orbital
magnetization in solids are collected in Section~\ref{sec:practical}. In
particular, we review options for taking the required $k$-space
derivatives in  Section~\ref{sec:k-space_derivatives} and how to perform
calculations using pseudopotentials in
Section~\ref{sec:pseudopotentials}. We further review a single $k$-point
formula (Section~\ref{sec:k-point}), a Wannier interpolation scheme
(Section~\ref{sec:Wannier_interpolation}), and DFT specific aspects
(Section~\ref{sec:DFT}). In Section~\ref{sec:FeCoNi} we summarize first
results for calculations of the orbital magnetization in Fe, Co, and Ni.
Finally, we review several applications of the orbital magnetization in
Section~\ref{sec:applications}. In particular, Sections~\ref{sec:NMR}
and \ref{sec:EPR} focus on showing how theories of nuclear magnetic
resonance (NMR) and electron paramagnetic resonance (EPR) can elegantly
be expressed as derivatives of the orbital magnetization. Other
important derivatives are reviewed in
Section~\ref{sec:other_derivatives}. We conclude with an outlook in
Section~\ref{sec:Conclusions}.

%%%%%%%%%%%%%%%%%%%%%%%%%%%%%%%%%%%%%%%%%%%%%%%%%%%%%%%%%%%%%%%%%%%%%%%%
\section{Importance of the Orbital Magnetization}
\label{sec:importance}
%%%%%%%%%%%%%%%%%%%%%%%%%%%%%%%%%%%%%%%%%%%%%%%%%%%%%%%%%%%%%%%%%%%%%%%%

As pointed out in the introduction, magnetism in solids originates from
two distinct sources, i.e.\ spin magnetic moments and orbital magnetic
moments. As such, the orbital magnetization is a crucial part of the
overall magnetization. The reason why it has not attracted more
attention in the past---and also the reason why the
condensed-matter/materials-science community ``survived'' without a
theory for it---is that in many common materials of everyday interest,
the orbital contribution is small compared to the spin contribution. In
Fe, Co, and Ni for example, it is only of the order of a few percent of
the total magnetization;\cite{Meyer_61} not surprising, its effects have
been neglected in the past. However, that is not to say that the orbital
magnetization is small in \emph{all} materials and in other systems it
can have a more important effect and even cancel the spin
magnetization.\cite{Qiao_2004, Gotsis_2003, Taylor_2002} Overall, the
orbital magnetization is important in-and-of-itself for a complete
description of magnetism in solids.

Furthermore, a wealth of applications are directly related to the
orbital magnetization. For example, a theory of solid-state NMR can
conveniently be defined in terms of a derivative of the orbital
magnetization,\cite{Thonhauser_2009a} discussed in detail in
Section~\ref{sec:NMR}, removing difficulties related to gauge-origin
choices of previous methods. In a similar sense, the EPR $g$-tensor
follows from a derivative of the orbital
magnetization,\cite{Ceresoli_2010a} as described in
Section~\ref{sec:EPR}. Other derivatives relate to properties such as
the magnetic susceptibility,  the orbital magnetoelectric coupling and
response,\cite{Malashevich_2010, Essin_2010, Essin_2009} the spin Hall
conductivity,\cite{Murakami_2006} and the identification of non-abelian
quantum Hall states.\cite{Cooper_2009}

The \emph{modern theory of orbital magnetization} is further important
because of its close connection to the \emph{modern theory of electric
polarization} in solids.\cite{King-Smith_93, Vanderbilt_93} In fact,
there is an underlying connection between these two theories through the
physics of the Berry phase,\cite{Berry_84} which has been discovered to
thread through many different fields of physics. This connection between
Berry's phase, the electric polarization, and the orbital magnetization
is thoroughly discussed in two other review articles.\cite{Xiao_2010,
Resta_2010}

Orbital magnetization permeates all dimensions---it plays an important
role in zero-dimensional,\cite{Blonski_2010, Lazarovits_2002,
Aldea_2003, Durr_1999, Bartolome_2008}
one-dimensional,\cite{Gambardella_2003, Hong_2004, Lazarovits_2004,
Gambardella_2002} two-dimensional,\cite{Oka_2009, Liu_2008, Xiao_2007,
Yao_2008, Faulhaber_2005, Oppen_2000, Meinel_2000, Wang_2007a} and
obviously three-dimensional systems.\cite{Qiao_2004, Gotsis_2003,
Taylor_2002, Wang_2007b, Cheng_2009, Todorova_2001, Brooks_1993,
Annett_2009, Braude_2006, Azimonte_2007, Lovesey_2002, Solovyev_1995}
The orbital magnetization is of interest in adatoms,\cite{Blonski_2010,
Lazarovits_2002} clusters and quantum dots,\cite{Aldea_2003, Durr_1999,
Bartolome_2008} and in metal- and nano-wires.\cite{Gambardella_2003,
Hong_2004, Lazarovits_2004, Gambardella_2002} It plays a role in
graphene\cite{Oka_2009, Liu_2008, Xiao_2007, Yao_2008} and the
two-dimensional electron gas.\cite{Faulhaber_2005, Oppen_2000,
Meinel_2000} It further is significant for certain
ferromagnets\cite{Wang_2007a} and antiferromagnets\cite{Wang_2007b}; in
some ferromagnets it can even compensate the spin magnetization,
resulting in zero net magnetization.\cite{Qiao_2004, Gotsis_2003,
Taylor_2002} The orbital magnetization is of importance in
semiconductors,\cite{Cheng_2009} metallic magnets,\cite{Todorova_2001}
itinerant magnets,\cite{Brooks_1993} and
superconductors.\cite{Annett_2009, Braude_2006} The orbital
magnetization can play an important part in certain
half-metals,\cite{Azimonte_2007} Mott insulators,\cite{Lovesey_2002} and
ordered alloys.\cite{Solovyev_1995} It influences the thermoelectric,
anomalous, and topological transport,\cite{Xiao_2007, Wang_2007a,
Xiao_2006} the magnetic susceptibility,\cite{Wang_2007a} as well as the
anisotropic magnetoresistance.\cite{Autes_2008} Finally, it also shows
interesting effects at interfaces,\cite{Koide_2001} in the quantum Hall
fluid,\cite{Geller_1995} in localized electrons,\cite{Wohlman_1995} and
in connection with the orbital polarization of itinerant
magnets.\cite{Solovyev_2005}

%%%%%%%%%%%%%%%%%%%%%%%%%%%%%%%%%%%%%%%%%%%%%%%%%%%%%%%%%%%%%%%%%%%%%%%%
\section{Theory of Orbital Magnetization in Solids}
\label{sec:theory}
%%%%%%%%%%%%%%%%%%%%%%%%%%%%%%%%%%%%%%%%%%%%%%%%%%%%%%%%%%%%%%%%%%%%%%%%

The theory of orbital magnetization was developed by two research groups
independently in 2005. Although their approaches are completely
different, they complement each other and the final expressions derived
within each approach---which we have indicated with boxes for easier
reference---all come to agreement. In the following sections we review
the theory presented in the corresponding original manuscripts. Note
that we are consistently using \emph{Gaussian units} throughout, which,
at times, deviates from the original manuscripts. However, before
showing details about the derivation of the theory of orbital
magnetization in solids, it is instructive to take a moment and consider
the original problem.

%%%%%%%%%%%%%%%%%%%%%%%%%%%%%%%%%%%%%%%%%%%%%%%%%%%%%%%%%%%%%%%%%%%%%%%%
\subsection{What's the problem?}
\label{sec:problem}

As pointed out in the introduction, the calculation of orbital
contributions to magnetism in atoms and molecules is rather
straightforward. But, what makes its calculation so difficult in
solids?

If we consider a finite system, the orbital contribution to the magnetic
moment can be expressed in terms of basic quantities. In a simplified
one-particle picture, defining the velocity operator $\vvec$ as the
commutator of the position operator $\rvec$ with the system's
Hamiltonian $H$,
\begin{equation}\label{equ:vvec}
\vvec = -\frac{i}{\hbar}[\rvec,H]\;,
\end{equation}
the total orbital moment in Gaussian units becomes
\begin{equation}\label{equ:moment}
\mvec\orb = -\frac{e}{2c}\sum_n f_n\;
\me{\psi_n}{\rvec\times\vvec}{\psi_n}\;.
\end{equation}
Here, $\ket{\psi_n}$ are the eigenstates of $H$, $f_n$ is the occupation
number, and $-e$ and $c$  are the electronic charge and the speed of
light, respectively. This is in parallel to the classically defined
moment caused by a current density $\Jvec(\rvec)$ in Gaussian
units,\cite{Jackson}
\begin{equation}
\mvec\orb = \frac{1}{2c}\int\dr\;\rvec\times\Jvec =
\frac{1}{2c}\int\dr\;\rvec\times\rho\vvec\;,
\end{equation}
where in the last step we have explicitly included the charge density
$\rho(\rvec)$. The magnetization $\Mvec\orb$ can then be defined as the
magnetic moment per unit volume $V$ as\footnote{The definition of a
magnetization, i.e.\ a magnetic moment per unit volume, is formally only
useful for periodic systems, but we use it here for finite systems
representing a fragment of a larger, crystalline system.}
\begin{equation}\label{equ:magnetization_finite}
\Mvec\orb = \frac{\mvec\orb}{V} = -\frac{e}{2cV}\sum_n f_n\;
\me{\psi_n}{\rvec\times\vvec}{\psi_n}\;.
\end{equation}
Equations~(\ref{equ:moment}) and (\ref{equ:magnetization_finite}) are
valid for all finite systems, even when very large. 

The essential difficulty for truly periodic systems becomes apparent
when taking the thermodynamic limit of
Eq.~(\ref{equ:magnetization_finite}). In such systems, the eigenstates
$\ket{\psi_n}$ become Bloch states $\ket{\psi_{n\kvec}}$ with Bloch wave
vector $\kvec$, in-between which the position operator is ill-defined
since Bloch states are extended. It is tempting to try to solve the
problem by simply changing to a more localized basis, but this alone
does not solve the problem, as we will show in
Section~\ref{sec:wannier}. Also, linear-response methods exist for
periodic systems that circumvent the problem in other ways, but they
only allow the calculation of magnetization \emph{changes}, not of the
magnetization itself.\cite{Mauri_96a, Mauri_96b, Pickard_02,
Sebastiani_01, Sebastiani_02}

One might also hope to solve the problem from the point of view of the 
local bulk current density $\Jvec(\rvec)$ and its relation to the
magnetization. However, while the eigenstates $\ket{\psi_{n\kvec}}$ can
be used to calculate $\Jvec(\rvec)$, it is often wrongly assumed that
this uniquely defines the magnetization. As Hirst has
emphasized,\cite{Hirst97} the knowledge of $\Jvec(\rvec)$---even in
principle---is insufficient to calculate the macroscopic orbital
magnetization $\Mvec\orb$. This can easily be seen from the following
argument. Consistent with Maxwell's equations we can define the orbital
magnetization density $\boldsymbol{\mathcal{M}}\orb(\rvec)$
as\cite{Hirst97}
\begin{equation}
c\,\nabla\times\boldsymbol{\mathcal{M}}\orb(\rvec) = \Jvec(\rvec)\;.
\end{equation}
However, $\boldsymbol{\mathcal{M}}\orb(\rvec)$ can simply be replaced by
another $\boldsymbol{\mathcal{M}}\orb'(\rvec)$, i.e.\
\begin{equation}
\boldsymbol{\mathcal{M}}\orb(\rvec) \Rightarrow
\boldsymbol{\mathcal{M}}\orb'(\rvec) =
\boldsymbol{\mathcal{M}}\orb(\rvec) +
\boldsymbol{\mathcal{M}}^0\orb +
\nabla\xi(\rvec)\;,
\end{equation}
which has in addition a constant shift $\boldsymbol{\mathcal{M}}^0\orb$
and even the gradient of an arbitrary function $\xi(\rvec)$, but
nevertheless produces the same local bulk current density
$\Jvec(\rvec)$.\cite{Todorova_2001} Hence,  the macroscopic orbital
magnetization $\Mvec\orb$ cannot be  uniquely defined  as the cell
average of $\boldsymbol{\mathcal{M}}\orb(\rvec)$. In this respect, the
theory of orbital magnetization is similar to the theory of electric
polarization, where knowledge of the density $\rho(\rvec)$---even in
principle---is insufficient to determine the electric polarization
$\Pvec$.\cite{King-Smith_93, Vanderbilt_93, polarization_3}

Since the calculation of the orbital magnetization in solids proves
difficult, a simple approximation is often used, referred to as
\emph{muffin-tin approximation}. Non-overlapping muffin-tin spheres can
be centered around the atoms in a solid. Within these spheres, which are
finite systems, the moment can then be calculated according to
Eq.~(\ref{equ:moment}). Often, the orbital magnetization indeed
originates from regions near the atom cores, making this approximation
good when the magnetization is mostly confined to the muffin-tin spheres
and the interstitial contribution can be neglected. This muffin-tin
approximation has for example been used in Refs.~[\citeonline{Wu_01,
Sharma_07}]. Now, having the full theory of orbital magnetization, we
can assess the accuracy of this approximation quantitatively, as done
for Fe, Co, and Ni in Section~\ref{sec:FeCoNi}.

%%%%%%%%%%%%%%%%%%%%%%%%%%%%%%%%%%%%%%%%%%%%%%%%%%%%%%%%%%%%%%%%%%%%%%%%
\subsection{Wannier function derivation of the orbital magnetization}
\label{sec:wannier}

For this first derivation of the orbital magnetization we limit
ourselves to insulating solids, described by a one-particle Hamiltonian
with broken time-reversal symmetry. We further restrict ourselves to
spinless electrons, one (completely occupied) bulk band, and vanishing
Chern invariant.\cite{Haldane_88, Ohgushi_00}. These restrictions are
non-essential and can easily be removed; a corresponding multi-band
case, metallic systems, and Chern insulators are discussed in
Section~\ref{sec:multi-band}. By assuming either a vanishing macroscopic
field or an integral number of flux quanta per unit cell, the Bloch wave
vector $\kvec$ is guaranteed to be a good quantum number. Breaking of
time-reversal symmetry in such situations can occur through spin-orbit
coupling to underlying ordered local moments.\cite{Haldane_88,
Ohgushi_00, Jungwirth_02, Murakami_03, Yao_04} We only focus on the main
concepts of the Wannier function derivative here and refer the reader to
the original manuscript in Ref.~[\citeonline{Thonhauser_05}] for further
details.

We start by considering a finite sample, which we then make larger and
larger until it becomes a periodic solid. The orbital magnetization of
the finite sample can be calculated using
Eq.~(\ref{equ:magnetization_finite}). However, as discussed in
Section~\ref{sec:problem}, the thermodynamic limit of this expression is
not well-defined. In parallel to the situation of the electric
polarization, it is tempting to switch from the delocalized Bloch-like
eigenstates $\ket{\psi_n}$  in Eq.~(\ref{equ:magnetization_finite}),
which cause the problem, to well-localized orthonormal orbitals
$\ket{\phi_n}$, simply by using the invariance of the trace. In the
thermodynamic limit, the $\ket{\phi_n}$ deep inside the sample can then
be associated with bulk Wannier functions $\ket{\Rvec}$, where $\Rvec$
is a lattice vector. Since Wannier functions are exponentially localized 
in insulators,\cite{Marzari_97, Brouder_2007, He_01} matrix elements of
the position operator $\rvec$ are now indeed well defined. It is at this
point that this derivation relies on the restrictions described above,
as Wannier functions in metals or Chern insulators are not exponentially
localized.\cite{Thonhauser_06}

Surprisingly, simply evaluating Eq.~(\ref{equ:magnetization_finite}) for
a solid in-between the bulk Wannier function $\ket{\Rvec}$ in the home
unit cell, i.e.\ $\ket{\home}$, does not result in the total orbital
magnetization.\cite{Resta_05} Instead, this leads to a contribution that
corresponds to the local- or self-circulation of a Wannier function
\begin{equation}\label{equ:M_LC_Wannier}
\Mvec\orb^\text{LC} = -\frac{e}{2cV_0}
\me{\home}{\rvec\times\vvec}{\home}\;,
\end{equation}
i.e.\ the circulation resulting from a Wannier function rotating around
its center. The volume $V_0$ corresponds to the unit-cell volume. Note
that the sum symbol from Eq.~(\ref{equ:magnetization_finite}) disappears
because we have limited ourselves to only one occupied bulk band,
resulting in only one occupied Wannier function. Interestingly, the
same approach in the case of the electric polarization $\Pvec$, i.e.\
replacing $\ket{\psi_n}$ in the analog of
Eq.~(\ref{equ:magnetization_finite}) with bulk Wannier functions
$\me{\psi_n}{\rvec}{\psi_n}\rightarrow\me{\home}{\rvec}{\home}$, solved
the problem completely.\cite{King-Smith_93, Vanderbilt_93, polarization_3}

The missing, second contribution to the magnetization arises from
itinerant surface contributions $\Mvec\orb^\text{IC}$, which---even in
periodic systems---have a remnant, fully contained in the bulk band
structure. To find this missing contribution, we need to find the
surface current caused by the itinerant circulation of the Wannier
functions. Note that  bulk Wannier functions of the sample, as they
correspond to fully occupied bulk bands, cannot carry any current and
thus cannot produce an itinerant contribution. Surface Wannier
functions, on the other hand, behave differently; they are not
eigenstates of the Hamiltonian and thus evolve over time. It is
important to note that this surface contribution is different from the
simple textbook picture where an array of current loops cancel inside a
domain and only the surface current remains. This itinerant contribution
originates from the fact that the surface Wannier functions or current
loops themselves move around the surface!

\begin{figure}[t]
{\bf a)}\quad\includegraphics[width=2in]{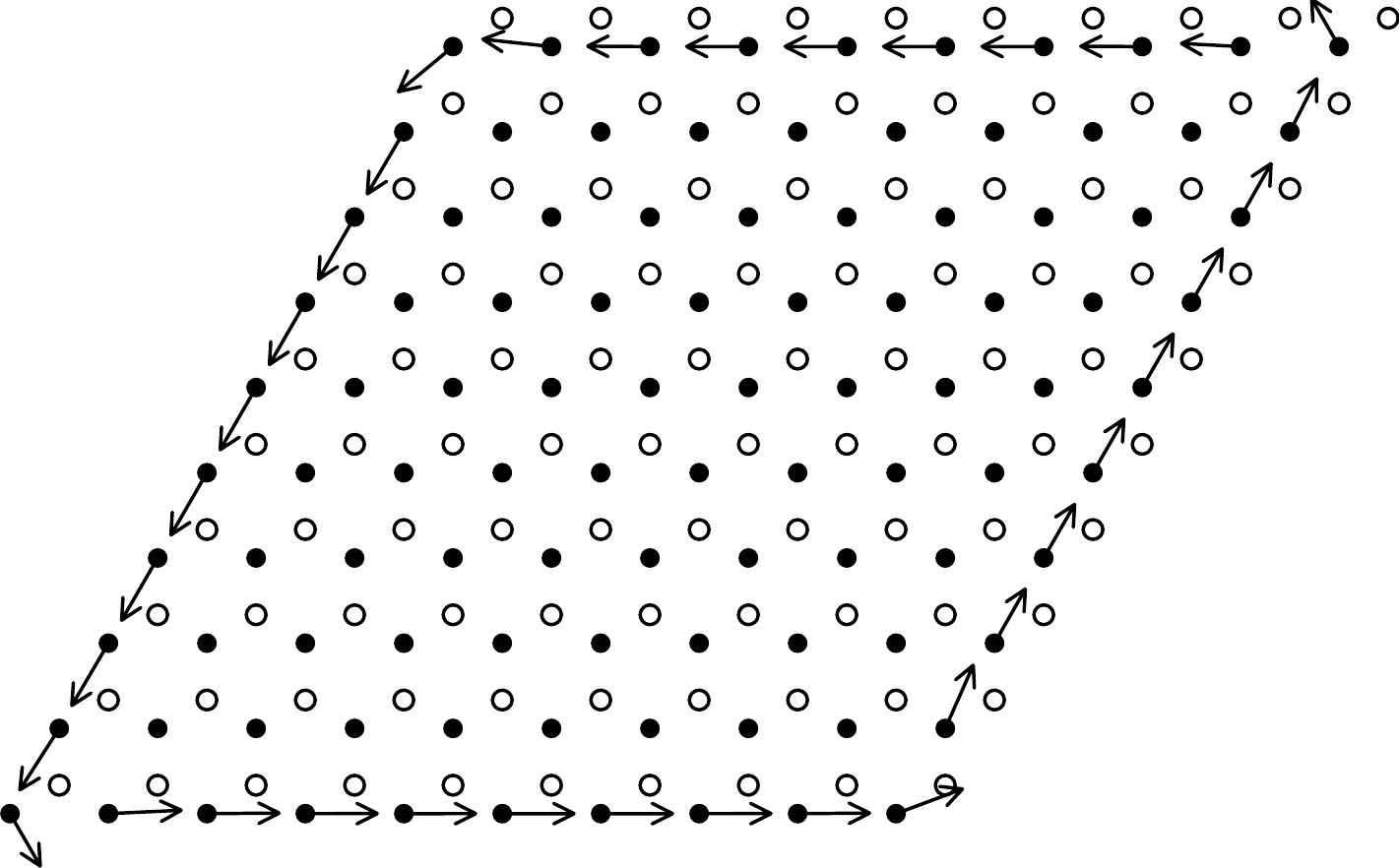}%
\hfill{\bf b)}\quad\includegraphics[width=2.4in]{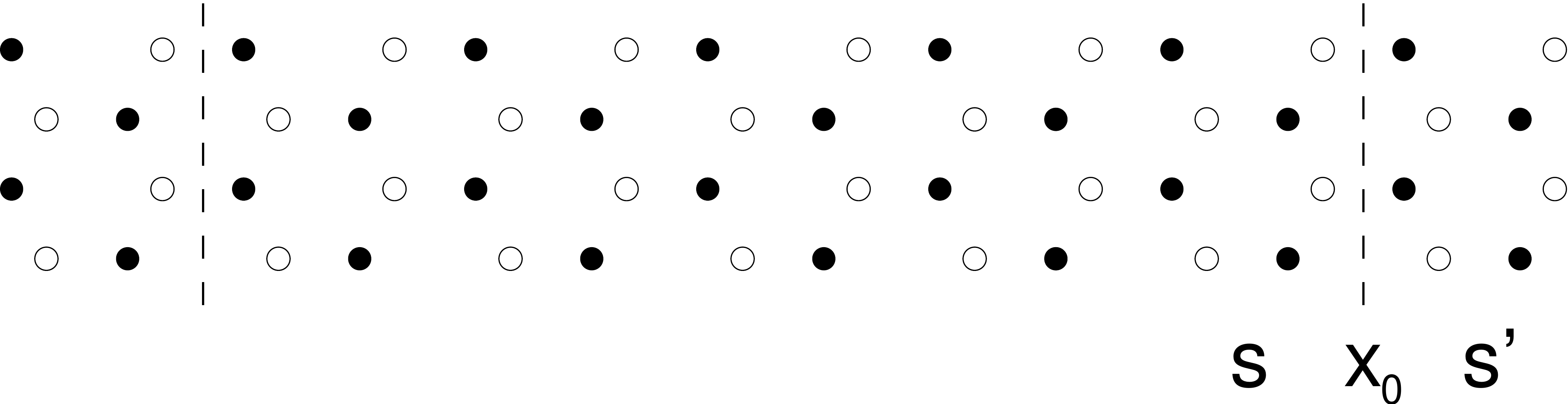}
\caption{\label{fig:haldane} {\bf a)} 10$\times$10 sample of the Haldane
model.\cite{Haldane_88} The currents $-e\me{\phi_n}{\vvec}{\phi_n}$
associated with each Wannier function are plotted using arrows at the
corresponding Wannier centers
$\bar{\rvec}_n=\me{\phi_n}{\rvec}{\phi_n}$. The itinerant current gives
rise to an additional contribution $\Mvec\orb^{\text{IC}}$ to the
orbital magnetization. {\bf b)}~Horizontal slice of the same materials
as in Fig.~\ref{fig:haldane}a, but infinitely long in the vertical
y-direction. (Reprinted with permission from
Ref.~[\citeonline{Thonhauser_05}]; \copyright\ 2005 American Physical
Society).}
\end{figure}

To show this fact, the authors of the original article in
Ref.~[\citeonline{Thonhauser_05}] used a simple, two-dimensional
tight-binding model.\cite{Haldane_88} A finite system consisting of
10$\times$10 unit cells of this hexagonal system is shown in
Fig.~\ref{fig:haldane}a. This figure also shows results for the
calculation of  the itinerant current
$-e\vvec_n=-e\me{\phi_n}{\vvec}{\phi_n}$ (indicated by arrows)
associated with each Wannier function, plotted at the Wannier centers 
$\bar{\rvec}_n=\me{\phi_n}{\rvec}{\phi_n}$. The current dies-off
exponentially into the bulk, as bulk Wannier functions do not carry any
current. The surface current obviously gives an additional contribution
to the magnetization according to
\begin{equation}
\Mvec\orb^{\text{IC}} = -\frac{e}{2cV}\sum_n\bar{\rvec}_n
\times\me{\phi_n}{\vvec}{\phi_n}\;.
\end{equation}

Below, we show how this itinerant surface contribution can be expressed
as a bulk property. To this end, we focus on the right edge of the
sample displayed in Fig.~\ref{fig:haldane}b. We use labels $s$ and $s'$
to denote Wannier functions to the left and right of the boundary at
$x_0$, respectively. The $y$-component of the current on the right edge
of the sample then becomes
\begin{equation}\label{equ:I_1}
I_y = -\frac{e}{\Delta l}{\sum_{s'}}'
\me{\phi_{s'}}{v_y}{\phi_{s'}} =
\frac{ie}{\hbar\Delta l}{\sum_{s'}}'
\me{\phi_{s'}}{[r_y,H]}{\phi_{s'}} =
-\frac{e}{\Delta l}\sum_s{\sum_{s'}}'
v_{\langle s,s'\rangle,y}\;,
\end{equation}
where we have used Eq.~(\ref{equ:vvec}) and inserted a complete set of
Wannier functions in-between $\rvec$ and $H$. In the last step we used
the notation $\vvec_{\langle
j,i\rangle}=(2/\hbar)\:\text{Im}\:\me{\phi_i}{\rvec}{\phi_j}
\me{\phi_j}{H}{\phi_i}$. Note that only Wannier functions that are
within the vertical extension $\Delta l$ of the strip contribute to the
primed sum. In principle, the first sum should run over all Wannier
functions $n$, but the ones where $n$ is within the subset of $s'$ do
not contribute to the overall current. Moving the boundary $x_0$ deep
into the bulk and using translational symmetry, the current in
Eq.~(\ref{equ:I_1}) becomes
\begin{equation}
I_y = -\frac{e}{\Delta l}\,{\sum_{R_x < x_0}}\;
{\sum_{R'_x > x_0}\!\!}' v_{\langle0,\Rvec'-\Rvec\rangle,y} =
-\frac{e}{2V_0}\,{\sum_{\Rvec}} R_x v_{\langle 0,\Rvec\rangle,y}\;,
\end{equation}
where we have further used the fact that there are exactly $(R'_x - R_x)
\Delta l/V_0$ terms in the sum with a given value of $\Rvec'-\Rvec$ if
$R_x'-R_x>0$ and zero otherwise. We divide by 2 in order to extend the
sum over all $\Rvec$. The $z$-component of the corresponding magnetic
moment then finally becomes
\begin{equation}
M_{\text{orb},z}^{\text{IC}} = -\frac{e}{2c\hbar V_0}\;
\text{Im}\;\sum_\Rvec\Big(
R_x\me{\Rvec}{r_y}{\home}\me{\home}{H}{\Rvec}-
R_y\me{\Rvec}{r_x}{\home}\me{\home}{H}{\Rvec}\Big)\;,
\end{equation}
or more generally
\begin{equation}\label{equ:M_IC_Wannier}
\Mvec_{\text{orb}}^{\text{IC}} = -\frac{e}{4cV_0}
\sum_\Rvec\Rvec\times\vvec_{\langle 0,\Rvec\rangle}\;,
\end{equation}
clearly expressing also the itinerant contribution
$\Mvec\orb^{\text{IC}}$ as a bulk property.

The final step is to transform the expressions for
$\Mvec\orb^{\text{LC}}$ and $\Mvec\orb^{\text{IC}}$ back to the Bloch
representation. With the definition of Wannier functions in terms of the
cell-periodic part of the Bloch functions
$\ket{u_\kvec}=e^{-i\kvec\cdot\rvec}\ket{\psi_\kvec}$, i.e.\
\begin{equation}
\ket{\Rvec} = \frac{V_0}{(2\pi)^3}\int\dk\;
e^{i\kvec\cdot(\rvec-\Rvec)}\,\ket{u_{\kvec}}\;,
\end{equation}
the main result becomes, after some straightforward but tedious
algebra,
\begin{equation}\label{equ:m_orb_wannier_single}
\fbox{$\displaystyle
\Mvec\orb^{} = \Mvec\orb^{\text{LC}} + \Mvec\orb^{\text{IC}} =
\frac{e}{2\hbar c} \,{\rm Im} \int \frac{\dk}{(2\pi)^3} \,
\me{\park u_\kvec}{\times (H_\kvec+E_\kvec)\,}{\park u_\kvec}\;.
$}\end{equation}
Here, we have used the shortcuts $\park\equiv\partial/\partial\kvec$
and  $H_\kvec\equiv e^{-i\kvec\cdot\rvec}\,H\,e^{i\kvec\cdot\rvec}$; the
$E_\kvec$ denote the corresponding eigenvalues. This expression is
reminiscent of a result found earlier for the special case of the
Hofstadter model.\cite{Gat_03b}. 

Equation~(\ref{equ:m_orb_wannier_single}) is the main result of this
section. It gives the orbital magnetization of a solid in terms of a
Brillouin-zone integral over well-understood quantities. Even in the
final expression, the two contributions corresponding to the internal
circulation of bulk Wannier functions (containing $H_\kvec$) and the net
currents carried by Wannier functions near the surface (containing
$E_\kvec$), are clearly distinguishable. Note that the surface
contribution can be written as 
\begin{equation}\label{equ:m_IC_Berry}
\Mvec\orb^{\text{IC}} = -\frac{e}{2\hbar c}
\int\frac{\dk}{(2\pi)^3} \, E_\kvec\;\boldsymbol{\Omega}_\kvec\;,
\end{equation}
where $\boldsymbol{\Omega}_\kvec = \nabla_\kvec\times\Avec_\kvec$ is the
Berry curvature and $\Avec_\kvec = i\me{u_\kvec}{\nabla_\kvec}{u_\kvec}$
the Berry connection---revealing the underlying relationship with the
theory of the Berry phase.\cite{Berry_84, Xiao_2010, Resta_2010} It is
interesting to note that both contributions to the orbital magnetization
in Eq.~(\ref{equ:m_orb_wannier_single}) are individually gauge
invariant\footnote{Here and in the following, we use the term
\emph{gauge invariance} to refer to the fact that a certain property is
independent of the choice of phase factors of the Bloch states. All
physical observables have to be gauge invariant in this sense.} and thus
are, in principle, separately observable.\cite{Ceresoli_2006} However,
this is only true for the simplified case of a single occupied bulk
band.

Although maximally localized Wannier functions were used to derive
Eq.~(\ref{equ:m_orb_wannier_single}), in the end the orbital
magnetization can be written in terms of Bloch states---already
suggesting that this expression might also be true for systems where
maximally localized Wannier functions cannot be constructed, such as
metals and Chern insulators.\cite{Thonhauser_06} In fact, in the next
sections we will see that this is indeed the case.

%%%%%%%%%%%%%%%%%%%%%%%%%%%%%%%%%%%%%%%%%%%%%%%%%%%%%%%%%%%%%%%%%%%%%%%%
\subsection{Multi-band derivation of the orbital magnetization\\
and extensions to metals and Chern insulators}
\label{sec:multi-band}

The derivation of the orbital magnetization in the previous section was
limited in that it only applied to insulators with vanishing Chern
invariant and one occupied bulk band. Not long after the original theory
appeared in Ref.~[\citeonline{Thonhauser_05}], these limitations were
removed in Ref.~[\citeonline{Ceresoli_2006}]. We review here the main
results of Ref.~[\citeonline{Ceresoli_2006}] and refer the reader to the
original manuscript for further details.

The derivative of a theory of orbital magnetization for multi-band
systems follows, in spirit, the outline of the previous section. We now
assume that there are several occupied bulk bands in the solid,
indicated by the band index $n$ in
$\ket{\psi_{n\kvec}}=e^{i\kvec\cdot\rvec}\ket{u_{n\kvec}}$, with
occupation $f_{n\kvec}$. This leads to the existence of several Wannier
functions in each unit cell labeled by $\ket{n\Rvec}$. It is easy to see
that, following the same arguments from Section~\ref{sec:wannier}, one
arrives at the same equations, with the only difference being that the
Wannier functions get an index and sums over bands occur. In summary,
Eqs.~(\ref{equ:M_LC_Wannier}) and (\ref{equ:M_IC_Wannier}) turn into
\begin{eqnarray}
\Mvec\orb^\text{LC}   &=& -\frac{e}{2cV_0}\sum_n\me{n\home}
                          {\rvec\times\vvec}{n\home}\;,\\
\Mvec\orb^{\text{IC}} &=& -\frac{e}{4cV_0}\sum_{nm\Rvec}\Rvec
                          \times\vvec_{\langle m\home,n\Rvec\rangle}\;,
\end{eqnarray}
which, after transformation back to Bloch states, becomes
\begin{eqnarray}\label{equ:m_orb_wannier_multi}
\Mvec\orb^{} &=& \Mvec\orb^{\text{LC}} +
                 \Mvec\orb^{\text{IC}} =\nonumber\\
             &=& \frac{e}{2\hbar c}\,{\rm Im}\sum_{n}\int
                 \frac{\dk}{(2\pi)^3}\;f_{n\kvec}\;
                 \me{\park u_{n\kvec}}{\times (H_\kvec+E_{n\kvec})\,}
                 {\park u_{n\kvec}}\;.
\end{eqnarray}

The main difference between Eq.~(\ref{equ:m_orb_wannier_multi}) and
Eq.~(\ref{equ:m_orb_wannier_single})---besides the band indices and sum
over bands---is the fact that in the single-band case both contributions
$\Mvec\orb^\text{LC}$ and $\Mvec_{\text{orb}}^{\text{IC}}$ by themselves
are gauge invariant, while in the multi-band case only their \emph{sum}
is. The corresponding proof goes beyond the scope of this review, but is
given in much detail in Ref.~[\citeonline{Ceresoli_2006}]. In the same
reference, it is also shown that terms in
Eq.~(\ref{equ:m_orb_wannier_multi}) can be regrouped such that the
overall magnetization is again written as a sum of two contributions,
whereas each contribution itself is now gauge invariant. This
relationship is further investigated in Ref.~[\citeonline{Souza_2008}],
where it is shown---through an elegant relation to the $f$-sum
rule---that the two contributions can be determined separately through a
combination of gyromagnetic and magneto-optical experiments.

Similar to the single-band case below Eq.~(\ref{equ:m_IC_Berry}), one
can define a multi-band Berry curvature as
\begin{equation}\label{equ:multi_band_omega}
\boldsymbol{\Omega}_{n\kvec} = i\me{\park u_{n\kvec}}{\times}
{\park u_{n\kvec}}\;,
\end{equation}
which in turn allows for a simple definition of the Chern invariant
\begin{equation}\label{equ:chern}
{\bf C}=\frac{1}{2\pi}\sum_n\int\dk\;\boldsymbol{\Omega}_{n\kvec}\;.
\end{equation}
It can now easily be seen that Eq.~(\ref{equ:m_orb_wannier_multi}) is
invariant with respect to shifts in the energy zero (as it must), as
long as the Chern invariant vanishes, which we have assumed so far. From
this, one can heuristically find an expression for the orbital
magnetization, even for cases in which the Chern invariant is not zero,
by simply enforcing the constant energy shift invariance again.  It
follows that a suitable extension of Eq.~(\ref{equ:m_orb_wannier_multi})
is
\begin{equation}\label{equ:m_orb_wannier_all}
\fbox{$\displaystyle
\Mvec\orb^{} = \frac{e}{2\hbar c} \,{\rm Im}\sum_{n}\int
\frac{\dk}{(2\pi)^3}\;f_{n\kvec}\;\me{\park u_{n\kvec}}
{\times (H_\kvec+E_{n\kvec}-2\mu)\,}{\park u_{n\kvec}}\;,
$}\end{equation}
where $\mu$ is the chemical potential. The factor of 2 in front of $\mu$
is necessary to compensate the shift of the energy zero, which affects
the Hamiltonian as well as the eigenvalues.

Equation~(\ref{equ:m_orb_wannier_all}) together with (\ref{equ:chern})
lead to
\begin{equation}
\frac{d\Mvec\orb}{d\mu} = \frac{e}{c\hbar(2\pi)^2}{\bf C}\;,
\end{equation}
for any insulator and $\mu$ in the gap, suggesting that the
magnetization changes linearly with $\mu$ if the Chern invariant is
non-zero. This peculiar result is a remnant of the surface states
present in Chern insulators\cite{Thonhauser_06} and has been proven
numerically.\cite{Ceresoli_2006}

Without proof, but based on numerical simulations, it can also be shown
that Eq.~(\ref{equ:m_orb_wannier_all}) holds for metallic 
systems,\cite{Ceresoli_2006} and is thus a generally valid expression
for all solids. Its correctness is further supported by alternative
derivations, which are presented in the following sections.

%%%%%%%%%%%%%%%%%%%%%%%%%%%%%%%%%%%%%%%%%%%%%%%%%%%%%%%%%%%%%%%%%%%%%%%%
\subsection{Semi-classical derivation of the orbital magnetization}
\label{sec:semi-classical}

Both of the previous sections derived a fully quantum-mechanical theory
of the orbital magnetization in solids based on a transformation to
well-localized Wannier functions to circumvent the problems of the
position operator in extended systems. At the same time the Wannier
function approach was developed, a separate research group---completely
independent---derived an identical expression of the orbital
magnetization through very elegant arguments concerning the
semi-classical equations of motion for Bloch electrons, as described in
detail in Ref.~[\citeonline{Xiao_2005}]. Its derivative, presented
below, and the derivations from the previous sections nicely complement
each other, each offering insight into certain aspects of the same
physics that the other cannot provide.

The derivation starts from the semi-classical equations of motion in
Gaussian units for an electron wave packet in band $n$, i.e.\
\begin{eqnarray}\label{equ:rdot}
\dot{\rvec}        &=& \frac{1}{\hbar}\frac{\partial E_{n\kvec}}
                       {\partial\kvec} - \dot{\kvec}\times
                       \boldsymbol{\Omega}_{n\kvec}\;,\\
\label{equ:kdot}
\hbar\,\dot{\kvec} &=& -e\,\Evec(\rvec)-\frac{e}{c}\,\dot{\rvec}
                       \times\Bvec(\rvec)\;,
\end{eqnarray}
where $\Evec(\rvec)$ and $\Bvec(\rvec)$ are the external electric and
magnetic field, respectively. For simplicity, in the following we will
assume spatially constant external fields. It is interesting to note
that the second term on the right hand side of Eq.~(\ref{equ:rdot}) is
often overlooked in elementary textbooks. This ``anomalous velocity''
containing the Berry curvature is important for properties such as the
anomalous Hall conductivity.\cite{Wang_06}

\begin{figure}[t]
\begin{center}
\includegraphics[width=0.35\textwidth]{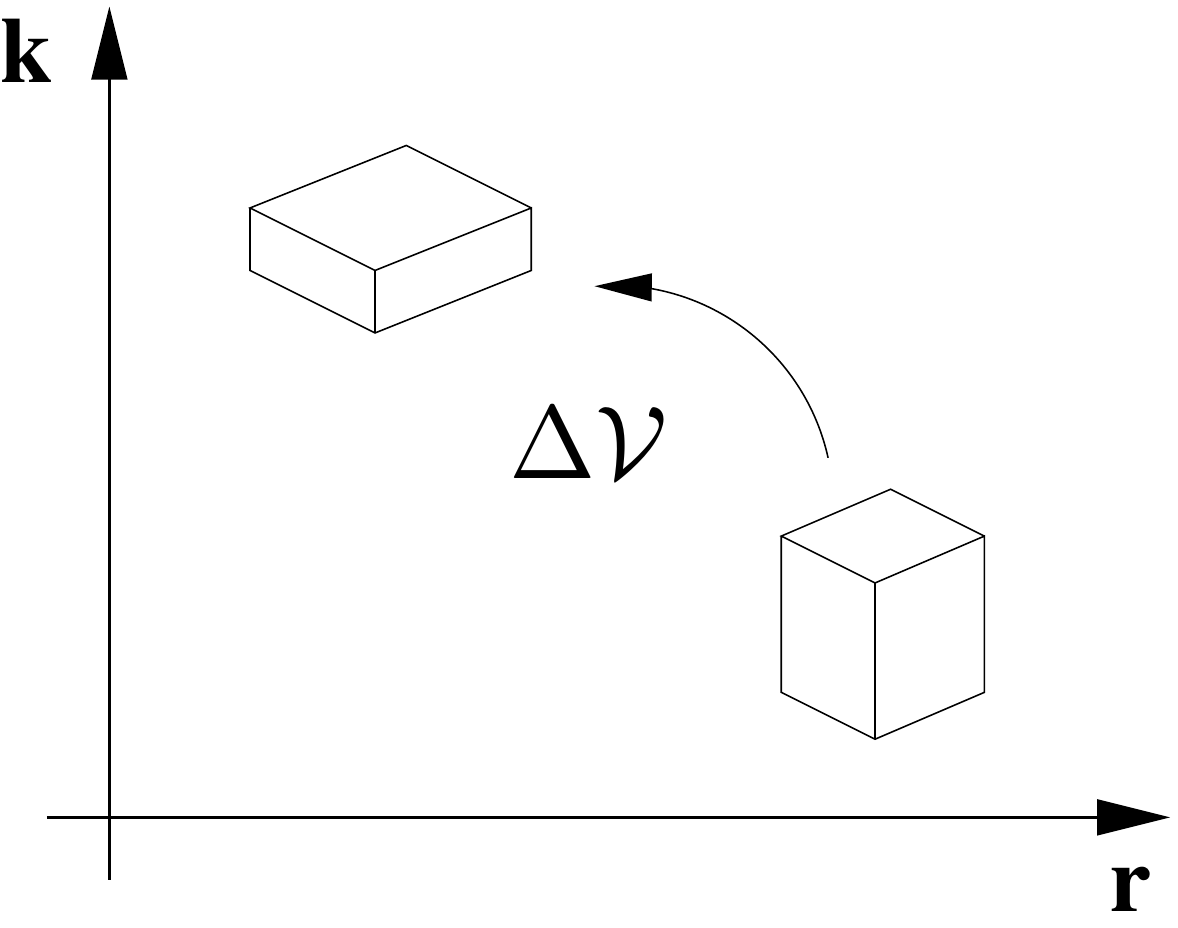}\\
\begin{minipage}{0.6\textwidth}
\caption{\label{fig:phase_space}Schematic depiction of the volume
element $\Delta\mathcal{V}=\Delta\rvec\Delta\kvec$ in phase space
$(\rvec,\kvec)$. The volume element might change its shape as a function
of time, but the volume itself is conserved according to Liouville's
theorem. (Figure modeled after Ref.~[\citeonline{Niu_05}]).}
\end{minipage}
\end{center}
\end{figure}

Closely linked to these semi-classical equations of motion is the
phase-space volume element $\Delta\mathcal{V} = \Delta\rvec\Delta\kvec$,
which Liouville's theorem predicts to be conserved, as is schematically
depicted in Fig.~\ref{fig:phase_space}. If we consider the equation of
motion for the volume element,\cite{Reichl_80}
$(1/\Delta\mathcal{V})(d\Delta \mathcal{V}/dt)=
\nabla_\rvec\cdot\dot{\rvec}+\nabla_\kvec\cdot\dot{\kvec}$, and insert
Eqs.~(\ref{equ:rdot}) and (\ref{equ:kdot}) we find for the time
evolution of the volume element
\begin{equation}
\Delta\mathcal{V} = \frac{\Delta\mathcal{V}_0}{1+e\Bvec\cdot
\boldsymbol{\Omega}_{n\kvec}/\hbar c}\;.
\end{equation}
Since $\boldsymbol{\Omega}_{n\kvec}$ depends on $\kvec$, it appears that
the volume element changes during the time evolution of the state
variables $(\rvec,\kvec)$, violating Liouville's theorem. The situation
can be remedied by introducing a modified density of states
\begin{equation}
D_{n\kvec} = \frac{1}{(2\pi)^3}(1+e\Bvec\cdot
\boldsymbol{\Omega}_{n\kvec}/\hbar c)\;,
\end{equation}
such that the number of states in the volume element $D_{n\kvec}\Delta
\mathcal{V}$ is constant again. Then, the expectation value of a
physical observable $\mathcal{O}$ can be written as
\begin{equation}\label{equ:phase_space}
\langle\mathcal{O}\rangle = \sum_n\int\dk\;f_{n\kvec}\;D_{n\kvec}\;
\me{\psi_{n\kvec}}{\,\mathcal{O}\,}{\psi_{n\kvec}}\;.
\end{equation}
For systems with time-reversal and inversion symmetry in the spatial
wave function the Berry curvature is zero and
Eq.~(\ref{equ:phase_space}) does not reveal any new physics. But, in
systems where the Berry curvature is non-zero such as  crystals with
broken time-reversal or inversion symmetry, Eq.~(\ref{equ:phase_space})
provides an elegant way to study Berry-phase effects.

In the original paper, Ref.~[\citeonline{Xiao_2005}], the authors give
several examples of how Eq.~(\ref{equ:phase_space}) can be used to
derive, in a simple manner, expressions for the electron density or the
anomalous Hall conductivity. In particular, it can be used to derive an
expression for the orbital magnetization. To that end, we describe an
electron semi-classically through a wave packet in Bloch band $n$. Such
a wave packet is found to rotate about its center of
mass,\cite{Sundaram_1999} leading to a magnetic moment\footnote{Note
that there is a factor of $-1/2$ missing in the original equation
published in Ref.~[\citeonline{Sundaram_1999}], as pointed out in
Ref.~[\citeonline{Xiao_2005}].}
\begin{equation}\label{equ:m_Niu}
\mvec_{n\kvec} = -\frac{ie}{2\hbar c}\me{\park u_{n\kvec}}
{\times\big(H_\kvec-E_{n\kvec}\big)}{\park u_{n\kvec}}\;.
\end{equation}
In a weak magnetic field $\Bvec$, the band structure
$E_{n\kvec}=\me{\psi_{n\kvec}}{H}{\psi_{n\kvec}}$ changes to $E_{n\kvec}
- \mvec_{n\kvec}\cdot\Bvec$. If we then express the total energy
according to Eq.~(\ref{equ:phase_space}) as
\begin{equation}
E = \sum_n\int\frac{\dk}{(2\pi)^3}\;f_{n\kvec}\;
\big(1+e\Bvec\cdot\boldsymbol{\Omega}_{n\kvec}/\hbar c\big)
\big(E_{n\kvec} - \mvec_{n\kvec}\cdot\Bvec\big)\;,
\end{equation}
the orbital magnetization follows by the negative derivative of the
energy with respect to $\Bvec$ in the zero field limit as
\begin{equation}\label{equ:semi_classical}
\fbox{$\displaystyle
\Mvec\orb^{} = \frac{e}{2\hbar c} \,{\rm Im}\sum_n
\int\frac{\dk}{(2\pi)^3}\;f_{n\kvec}\;
\me{\park u_{n\kvec}}{\times (H_\kvec+E_{n\kvec}-2\mu)\,}
{\park u_{n\kvec}}\;,
$}\end{equation}
in perfect agreement with Eqs.~(\ref{equ:m_orb_wannier_single}) and
(\ref{equ:m_orb_wannier_all}).

Since this semi-classical derivation does not rely on the localization
properties of Wannier functions, it is valid for all insulators and
metals alike, further supporting the heuristic arguments from
Section~\ref{sec:multi-band}.

%%%%%%%%%%%%%%%%%%%%%%%%%%%%%%%%%%%%%%%%%%%%%%%%%%%%%%%%%%%%%%%%%%%%%%%%
\subsection{Finite-temperature derivation  of the orbital magnetization
and generalization to interacting systems}
\label{sec:finite-temperature}

The derivatives of the orbital magnetization formula from the previous
sections had limitations in that they were either only valid for
insulators or they were based on semi-classical arguments. Here, we
review a completely general derivation of the same expression, which is
free of these limitations and was first presented in
Ref.~[\citeonline{Shi_2007}].

The starting point for this derivation is the thermodynamic definition
of the orbital magnetization
\begin{equation}
\Mvec\orb = -\frac{1}{V_0}\bigg(\frac{\partial\Omega_G}
{\partial\Bvec}\bigg)_{T,\mu}\;,
\end{equation}
where $\Omega_G=E-TS-\mu N$ is the grand canonical potential, $\Bvec$ is
the magnetic field, and $V_0$ is the unit-cell volume. The derivative is
to be taken at constant temperature $T$ and chemical potential $\mu$. It
is convenient to first focus on the related quantity
\begin{equation}
\widetilde{\Mvec}\orb = -\frac{1}{V_0} \bigg(\frac{\partial K}
{\partial\Bvec} \bigg)_{T,\mu}\;,
\end{equation}
with $K=E-\mu N$. The original orbital magnetization can then be
recovered by integration of
\begin{equation}\label{equ:integration}
\widetilde{\Mvec}\orb = \frac{\partial}{\partial\beta}\,
\beta\,\Mvec\orb\;,
\end{equation}
where $\beta=1/kT$.

Next, we apply an external, slowly varying field $\Bvec(\rvec)$ with
vector potential $\Avec(\rvec)$ and wave vector $q$, pointing in the
$\hat{\bf z}$-direction according to
\begin{equation}
\Bvec(\rvec) = B \cos(qy)\,\hat{\bf z}\;,\qquad\Avec(\rvec) =
-B\frac{\sin qy}{q}\,\hat{\bf x}\;.
\end{equation}
At the end of this derivation, we will take the limit $q\to 0$. If we
consider the original system to be described by the Hamiltonian $H_0$
with Bloch eigenfunctions $\psi_{n\kvec}$ and eigenvalues $E_{n\kvec}$,
we can treat the magnetic field as a perturbation, e.g.\ $H=H_0+H_B$,
with
\begin{equation}
H_B = \frac{e}{2c}\big[\vvec\cdot\Avec(\rvec) +
\Avec(\rvec)\cdot\vvec\big]\;.
\end{equation}

We can then define the grand-canonical ensemble energy density as
\begin{equation}
K(\rvec) = \sum_n\int\frac{\dk}{(2\pi)^3}\;
f_{n\kvec}\;{\rm Re}\;\Big[\psi_{n\kvec}^*(\rvec)\,
K\,\psi_{n\kvec}(\rvec)\Big]\;,
\end{equation}
where $K$ is now the operator $K=H-\mu N$ and $K_0=H_0-\mu N$.
Perturbation theory yields in first order a change in $K$ according to
\begin{eqnarray}\label{equ:K_perturbation_1}
\Delta K(\rvec) &=& \sum_n\int\frac{\dk}{(2\pi)^3}\;{\rm Re}\;
                    \Big[\Delta f_{n\kvec}^{}\,\psi_{n\kvec}^*\,K_0^{}\,
                    \psi_{n\kvec}^{} + 
                    f_{n\kvec}^{}\,\psi_{n\kvec}^*\,H_B^{}\,
                    \psi_{n\kvec}^{}\nonumber\\
                &&  {}+ f_{n\kvec}^{} \Big(
                    \Delta\psi_{n\kvec}^*\,K_0^{}\,\psi_{n\kvec}^{} +
                    \psi_{n\kvec}^*\,K_0^{}\,\Delta\psi_{n\kvec}^{}
                    \Big)\Big]\;.
\end{eqnarray}
However, in linear order in $B$ the change in $K$ can also be written as
\begin{equation}\label{equ:K_perturbation_2}
\Delta K(\rvec) = -\widetilde{\Mvec}\orb\cdot\Bvec(\rvec)\;,
\end{equation}
such that, through comparison of Eqs.~(\ref{equ:K_perturbation_1}) and
(\ref{equ:K_perturbation_2}), one can find an explicit expression for
$\widetilde{\Mvec}\orb$. Note that the first two terms of
Eq.~(\ref{equ:K_perturbation_1}) do not contribute to
$\widetilde{\Mvec}\orb$.\cite{Shi_2007} Inserting the first order change
of the wave function $\psi_{n\kvec}$ in Eq.~(\ref{equ:K_perturbation_1})
and taking the limit $q\to 0$, one finds after some straightforward but
tedious algebra\footnote{Note that the cross product $\times$ is missing
in the original expression in Eq.~(13) of Ref.~[\citeonline{Shi_2007}].}
\begin{eqnarray}\label{equ:final_M_tilde}
\widetilde{\Mvec}\orb &=& \frac{e}{2\hbar c} \,{\rm Im} \sum_n\int
                          \frac{\dk}{(2\pi)^3}\;\Big[\;
                          f_{n\kvec}\;\me{\park u_{n\kvec}}
                          {\times(H_{\kvec}+E_{n\kvec}-2\mu)}
                          {\park u_{n\kvec}}\nonumber\\
                      &&  {}-{\textstyle\frac{\partial f_{n\kvec}}
                          {\partial E_{n\kvec}}}(E_{n\kvec}-\mu)
                          \me{\park u_{n\kvec}}
                          {\times(E_{n\kvec}-H_{\kvec})}
                          {\park u_{n\kvec}}\Big]\;.
\end{eqnarray}

For $T=0$, the derivative $\frac{\partial f_{n\kvec}}{\partial
E_{n\kvec}}$ becomes a $\delta$-function of $(E_{n\kvec}-\mu)$ and the
second term in Eq.~(\ref{equ:final_M_tilde}) vanishes, such that the
remainder is in perfect agreement with
Eqs.~(\ref{equ:m_orb_wannier_single}), (\ref{equ:m_orb_wannier_all}),
and (\ref{equ:semi_classical}) from the previous sections. A finite
temperature expression can be simply derived through integration of
Eq.~(\ref{equ:integration}) and yields
\begin{equation}\label{equ:M_orb_finite}
\fbox{$\displaystyle
\Mvec\orb = \sum_n\int\frac{\dk}{(2\pi)^3}\;\Big[
f_{n\kvec}\,\mvec_{n\kvec} + \frac{e}{c\hbar\beta}
\boldsymbol{\Omega}_{n\kvec}
\ln\big[1+e^{-\beta(E_{n\kvec}-\mu)}\big]\Big]\;,
$}\end{equation}
where the Berry curvature $\boldsymbol{\Omega}_{n\kvec}$ and
$\mvec_{n\kvec}$ are defined in Eqs.~(\ref{equ:multi_band_omega}) and
(\ref{equ:m_Niu}), respectively.

In Ref.~[\citeonline{Shi_2007}] the authors go on to show that 
Eq.~({\ref{equ:M_orb_finite}}), besides being valid for insulators with
or without Chern invariants and metals at zero or finite temperatures,
is also valid for weak and strong magnetic fields. It further is
applicable to interacting systems, provided that the single particle
energies and wave functions are computed using current and spin density
functional theory (CSDFT).\cite{Vignale_88} As such,
Eq.~({\ref{equ:M_orb_finite}}) presents the most general derivation and
concludes our review of derivations for the theory of orbital
magnetization.

%%%%%%%%%%%%%%%%%%%%%%%%%%%%%%%%%%%%%%%%%%%%%%%%%%%%%%%%%%%%%%%%%%%%%%%%
\section{Practical Aspects of Calculating the Orbital Magnetization}
\label{sec:practical}
%%%%%%%%%%%%%%%%%%%%%%%%%%%%%%%%%%%%%%%%%%%%%%%%%%%%%%%%%%%%%%%%%%%%%%%%

%%%%%%%%%%%%%%%%%%%%%%%%%%%%%%%%%%%%%%%%%%%%%%%%%%%%%%%%%%%%%%%%%%%%%%%%
\subsection{k-space derivatives}
\label{sec:k-space_derivatives}

The expressions for the orbital magnetization in
Eqs.~(\ref{equ:m_orb_wannier_single}), (\ref{equ:m_orb_wannier_all}),
(\ref{equ:semi_classical}), and (\ref{equ:M_orb_finite}) contain the
$k$-space gradient of the cell-periodic part of the Bloch function,
i.e.\ $\ket{\park u_{n\kvec}}$ with $\park\equiv\partial/\partial\kvec$,
the evaluation of which is not necessarily trivial. In the following, we
review three practical options to compute this derivative. 

The first option can be considered \emph{analytical} and is based on
standard perturbation theory. The change of the states in first order to
a $\kvec\cdot{\bf p}$ perturbation can be expressed as a sum over
states, i.e.\
\begin{equation}\label{equ:sum_over_states}
\ket{\park u_{n\kvec}} = \sum_{m\ne n}\ket{u_{m\kvec}}
\frac{\me{u_{m\kvec}}{\vvec}{u_{n\kvec}}}{E_{n\kvec}-E_{m\kvec}}\;,
\end{equation}
where the velocity operator $\vvec$ is defined in Eq.~(\ref{equ:vvec}).
The disadvantage of this expression is that \emph{all} states have to be
included, even unoccupied ones. However, often first-principles computer
codes---to save compute time---do not calculate many unoccupied states.
Also, when plane waves are used, the sum converges very slowly and
becomes unpractical. Equation~(\ref{equ:sum_over_states}) is, however,
useful in the context of localized basis sets or tight-binding models,
where the number of states is small. In particular, this method has been
used to calculate the results presented in Section~\ref{sec:k-point}.

The second option is \emph{numerical} in nature, making use of the
covariant derivative.\cite{Sai_02, Souza_04} The numerical derivative is
not trivial, as can be seen from the following example: let $\qvec$ be a
vector that connects $\kvec$ with a nearby point in  direction $\alpha$.
In principle, we would then like to approximate the desired derivative
as the finite difference
\begin{equation}\label{equ:phase_issue}
\ket{\partial_\alpha u_{n\kvec}} \approx \frac{1}{2|\qvec|}\Big(
\ket{u_{n\kvec+\qvec}}-\ket{u_{n\kvec-\qvec}}\Big)\;,
\end{equation}
and the expression becomes exact in the limit of $|\qvec|\to 0$.
However, numerically this is not a well-defined quantity. While the
states $\ket{u_{n\kvec+\qvec}}$ and $\ket{u_{n\kvec-\qvec}}$ can easily
be obtained through diagonalization of the Hamiltonian at these points,
they in general will have arbitrary and unrelated overall phase factors,
depending on the diagonalization routine used. Thus, simply numerically
subtracting the states yields unpredictable results.

The problem of the phase can be circumvented by replacing the  normal
derivative with the covariant derivative\cite{Sai_02, Souza_04}
\begin{equation}\label{equ:covariant_derivative}
\ket{\partial_\alpha u_{n\kvec}}\to
\ket{\tilde{\partial}_\alpha u_{n\kvec}} =
\mathcal{Q}_\kvec\ket{\partial_\alpha u_{n\kvec}}\;.
\end{equation}
Here, $\mathcal{Q}_\kvec$ projects onto the empty states at point
$\kvec$ and is defined as the complement of $\mathcal{P}_\kvec$, i.e.\
$\mathcal{Q}_\kvec=1-\mathcal{P}_\kvec$, which projects onto the
occupied manifold
$\mathcal{P}_\kvec=\sum_n^{\text{occ}}\ket{u_{n\kvec}}\bra{u_{n\kvec}}$.
Note that the definition of a derivative on such a manifold is not
unique and the covariant derivative is a choice; other choices exist,
which also have components in the occupied states. We can then define
dual states $\ket{\tilde{u}_{n\kvec+\qvec}}$ as a linear combination of
the occupied states $\ket{u_{m\kvec+\qvec}}$ at  point $\kvec+\qvec$ as 
\begin{equation}\label{equ:dual_states}
\ket{\tilde{u}_{n\kvec+\qvec}} = \sum_m
(S^{-1}_{\kvec,\kvec+\qvec})_{mn}\ket{u_{m\kvec+\qvec}}\;,
\end{equation}
where $S$ is the overlap matrix
\begin{equation}
(S_{\kvec,\kvec+\qvec})_{nm} =
\langle u_{n\kvec}|u_{m\kvec+\qvec}\rangle\;.
\end{equation}
For sufficiently small $\qvec$, $S$ becomes nearly diagonal, containing
only the arbitrary overall phase factors of $\bra{u_{n\kvec}}$ and
$\ket{u_{n\kvec+\qvec}}$ in its diagonal elements. The covariant
derivative can then simply be written as
\begin{equation}
\ket{\tilde{\partial}_\alpha u_{n\kvec}} \approx
\frac{1}{2|\qvec|} \Big(\ket{\tilde{u}_{n\kvec+\qvec}} -
\ket{\tilde{u}_{n\kvec-\qvec}}\Big)\;,
\end{equation}
which is manifestly gauge independent in the sense that it does not
depend on the choice of phases for the $\ket{u_{n\kvec\pm\qvec}}$. It
thus resolves the phase issue of Eq.~(\ref{equ:phase_issue}) and is
numerically accessible. Note that the covariant derivative
$\ket{\tilde{\partial}_\alpha u_{n\kvec}}$ still has the same overall
phase factor as $\ket{u_{n\kvec}}$, which then cancels when evaluating
matrix elements such as the orbital magnetization. The effect that the
overlap matrix plays in canceling the arbitrary overall phase can be
seen more clearly if only one band is considered, as is done explicitly
in Eq.~(26) of Ref.~[\citeonline{Sai_02}].

The dual states enjoy the property of being orthogonal to the original
states
\begin{equation}\label{equ:ortho_dual}
\langle u_{n\kvec}|\tilde{u}_{m\kvec\pm\qvec}\rangle = \delta_{nm}\;,
\end{equation}
from which follows immediately that
\begin{equation}
\langle u_{n\kvec} \ket{\tilde{\partial}_\alpha u_{m\kvec}} =
\frac{1}{2|\qvec|}(\delta_{nm}-\delta_{nm})=0\;.
\end{equation}
In other words, the covariant derivative is orthogonal to the occupied
states, as evident by the projector $\mathcal{Q}_\kvec$ in
Eq.~(\ref{equ:covariant_derivative}).

The covariant derivative requires that the number of occupied states at
$\kvec$ and $\kvec\pm\qvec$ is the same, i.e.\ the material has to be an
insulator.  This is a remnant of the fact that it was first developed in
the context of the electric polarization,\cite{Sai_02, Souza_04} which
is only defined in insulators. However, it is conceivable that the
formalism of the covariant derivative can be extended to metallic
systems. 

The third option for calculating $k$-space derivatives is based on
linear-response theory. If we multiply Eq.~(\ref{equ:sum_over_states})
from the left with $H_\kvec-E_{n\kvec}$, we find
\begin{equation}
\big(H_\kvec-E_{n\kvec}\big)\ket{\park u_{n\kvec}} =
-\mathcal{Q}_{n\kvec}\vvec\ket{u_{n\kvec}}\;,
\end{equation}
where $\mathcal{Q}_{n\kvec}=1-\ket{u_{n\kvec}}\bra{u_{n\kvec}}$. This is
a Sternheimer equation, similar to Eq.~(25) in
Ref.~[\citeonline{Baroni_01}]. The advantage of this expression is that
the expensive sum over all states vanishes and the right hand side can
be evaluated with only knowledge of the occupied states. The equation
then forms a linear system, which can easily be inverted to find
$\ket{\park u_{n\kvec}}$ by iterative algorithms, such as a
conjugate-gradient approach. This linear-response approach can be used
to calculate the orbital magnetization in real, metallic systems and, in
particular, has been applied to calculate the orbital magnetization in
Fe, Co, and Ni, discussed in Section~\ref{sec:FeCoNi}.

%%%%%%%%%%%%%%%%%%%%%%%%%%%%%%%%%%%%%%%%%%%%%%%%%%%%%%%%%%%%%%%%%%%%%%%%
\subsection{Calculating the orbital magnetization\\
in a pseudopotential context}
\label{sec:pseudopotentials}

Many first-principles electronic-structure codes use plane waves as
basis functions for their many conveniences. However, to allow for  a
reasonable basis-set size the true Coulombic potential of the ions is
usually replaced by an effective potential, i.e.\ pseudopotential. Since
the expressions for the orbital magnetization in
Eqs.~(\ref{equ:m_orb_wannier_single}), (\ref{equ:m_orb_wannier_all}),
(\ref{equ:semi_classical}), and (\ref{equ:M_orb_finite}) are strictly
speaking only valid in an all-electron picture, they have to be modified
to be used within a pseudopotential framework. The corresponding theory
has been recently developed for the special case of norm-conserving
pseudopotentials in Ref.~[\citeonline{Ceresoli_2010a}], and we present
here a summary thereof.

In the pseudopotential formalism, the pseudo Hamiltonians $\bar{H}$ and
$\bar{H}_\kvec$ act on the pseudo wave functions
$\ket{\bar{\psi}_{n\kvec}}$ and $\ket{\bar{u}_{n\kvec}}$ with
eigenvalues $\bar{E}_{n\kvec}$. As a first approximation, either of the
expressions in Eqs.~(\ref{equ:m_orb_wannier_single}),
(\ref{equ:m_orb_wannier_all}), (\ref{equ:semi_classical}), or 
(\ref{equ:M_orb_finite}) can simply be evaluated using the pseudo
quantities $\bar{H}_\kvec$, $\ket{\bar{u}_{n\kvec}}$, and
$\bar{E}_{n\kvec}$ instead of the all-electron ones $H$,
$\ket{u_{n\kvec}}$, and $E_{n\kvec}$. This results---as we shall see in
Section~\ref{sec:FeCoNi}---in only a small error in the orbital
magnetization, on the order of a few percent.

The true wave functions can be recovered from the pseudo ones by means
of the projector augmented waves (PAW) transformation
$\ket{\psi_{n\kvec}} =
\mathcal{T}\ket{\bar{\psi}_{n\kvec}}$.\cite{Bloechl_94} In the presence
of magnetic fields, the PAW transformation has to be modified, resulting
in a gauge including projector augmented waves (GIPAW)\cite{GIPAW}
formalism with the corresponding transformation $\ket{\psi_{n\kvec}} =
\mathcal{T}_B\ket{\bar{\psi}_{n\kvec}}$.\cite{Pickard_01}  The pseudo
Hamiltonian then is given through  $\bar{H}=\mathcal{T}_B^\dag H
\mathcal{T}_B^{}$. If we define the orbital  magnetization as the
derivative of the total energy,\footnote{Here, we are not including the
trivial spin-Zeeman term in the Hamiltonian; if we did, the energy
derivative would correspond to the total magnetization.} it follows that
\begin{eqnarray}
\Mvec\orb = -\frac{1}{V_0}
       \frac{\partial E_\text{tot}}{\partial\Bvec}\bigg|_{B=0}
   &=& -\frac{1}{V_0}\sum_n\int \frac{\dk}{(2\pi)^3} \,f_{n\kvec}\,
       \frac{\partial}{\partial\Bvec}\me{\psi_{n\kvec}}
       {H}{\psi_{n\kvec}}_{\raisebox{-0.5ex}{$_{B=0}$}}\nonumber\;,\\
   &=& -\frac{1}{V_0}\sum_n\int \frac{\dk}{(2\pi)^3} \,f_{n\kvec}\,
       \frac{\partial}{\partial\Bvec}\me{\bar{\psi}_{n\kvec}}
       {\bar{H}}{\bar{\psi}_{n\kvec}}_{\raisebox{-0.5ex}{$_{B=0}$}}\;,
       \nonumber\\
   &=& -\frac{1}{V_0}\sum_n\int \frac{\dk}{(2\pi)^3} \,f_{n\kvec}\,
       \me{\bar{\psi}_{n\kvec}}{\frac{\partial\bar{H}}{\partial\Bvec}}
       {\bar{\psi}_{n\kvec}}_{\raisebox{-0.5ex}{$_{B=0}$}}\;,
\end{eqnarray}
where we have used the Hellmann-Feynman theorem in the last step.
Explicitly calculating the pseudo Hamiltonian through the GIPAW
transformation, taking the $\Bvec$ derivative, and finally taking the
$\Bvec=0$ limit yields
\begin{equation}\label{equ:M_orb_pseudo_tot}
\Mvec\orb^{} = \Mvec_{\text{orb}}^{\text{bare}} +
               \Delta\Mvec_{\text{orb}}^{\text{bare}} +
               \Delta\Mvec_{\text{orb}}^{\text{para}} +
               \Delta\Mvec_{\text{orb}}^{\text{dia}}\;,
\end{equation}
with\footnote{Note that there is a factor of $-1/V_0$ missing in
Eqs.~(10)--(13) in Ref.~[\citeonline{Ceresoli_2010a}].}
\begin{eqnarray}\label{equ:M_orb_pseudo_dbare}
\Delta\Mvec_{\text{orb}}^{\text{bare}}
   &=& \frac{e}{2\hbar cV_0}\sum_{n\Rvec}
       \int \frac{\dk}{(2\pi)^3} \,f_{n\kvec}\,
       \me{\bar{u}_{n\kvec}}{(\Rvec-\rvec)\times i\;
       \big[\rvec-\Rvec,V_\Rvec^{\text{NL}}\big]\,}
       {\bar{u}_{n\kvec}}\;,\\
\Delta\Mvec_{\text{orb}}^{\text{para}}
   &=& \frac{g'e}{16m_e^2c^3V_0}\sum_{n\Rvec}
       \int \frac{\dk}{(2\pi)^3} \,f_{n\kvec}\,
       \me{\bar{u}_{n\kvec}}{(\Rvec-\rvec)\times i\;
       \big[\rvec-\Rvec,F_\Rvec^{\text{NL}}\big]\,}
       {\bar{u}_{n\kvec}}\;,\quad\\
\label{equ:M_orb_pseudo_ddia}
\Delta\Mvec_{\text{orb}}^{\text{dia}}
   &=& -\frac{g'\hbar e}{16m_e^2c^3V_0}\sum_{n\Rvec}
       \int \frac{\dk}{(2\pi)^3} \,f_{n\kvec}\,
       \me{\bar{u}_{n\kvec}}{{\bf E}_\Rvec^{\text{NL}}}
       {\bar{u}_{n\kvec}}\;.
\end{eqnarray}
Here, $g'\approx 2.004\,639$,\cite{Schreckenbach_97} $m_e$ is the
electron mass, $V_\Rvec^{\text{NL}}$ is the non-local part of the
pseudopotential in separable form, and $F_\Rvec^{\text{NL}}$ and ${\bf
E}_\Rvec^{\text{NL}}$ are the separable non-local GIPAW  projectors at
atomic site $\Rvec$. If we denote the GIPAW projector as
$\ket{\bar{p}_{n\Rvec}}$ and the all-electron and pseudo
partial waves as $\ket{\phi_{n\Rvec}}$ and $\ket{\bar{\phi}_{n\Rvec}}$,
respectively, the latter terms can be written as
\begin{eqnarray}
F_\Rvec^{\text{NL}}
   &=& \sum_{nm}\ket{\bar{p}_{n\Rvec}}\;
       \boldsymbol{\sigma}\cdot\Big[\me{\phi_{n\Rvec}}
       {\nabla V\times{\bf p}}{\phi_{m\Rvec}} -
       \me{\bar{\phi}_{n\Rvec}}{\nabla\bar{V}\times{\bf p}}
       {\bar{\phi}_{m\Rvec}}\Big]\;\bra{\bar{p}_{m\Rvec}}\;,\\
{\bf E}_\Rvec^{\text{NL}}
   &=& \sum_{nm}\ket{\bar{p}_{n\Rvec}}\;
       \Big[\me{\phi_{n\Rvec}}{\rvec\times(\boldsymbol{\sigma}
       \times\nabla V)}{\phi_{m\Rvec}} -
       \me{\bar{\phi}_{n\Rvec}}{\rvec\times
       (\boldsymbol{\sigma}\times\nabla\bar{V})}
       {\bar{\phi}_{m\Rvec}}\Big]\;\bra{\bar{p}_{m\Rvec}}\;\nonumber,
\end{eqnarray}
where $\boldsymbol{\sigma}$ are the Pauli matrices and $V$ and $\bar{V}$
are the screened atomic all-electron potential and the local
pseudopotential, respectively.

The $ \Mvec_{\text{orb}}^{\text{bare}}$ term on the right hand side of
Eq.~(\ref{equ:M_orb_pseudo_tot}) is indeed the earlier derived
expression for the orbital
magnetization---Eqs.~(\ref{equ:m_orb_wannier_single}),
(\ref{equ:m_orb_wannier_all}), (\ref{equ:semi_classical})---simply
evaluated for the pseudo Hamiltonian $\bar{H}_\kvec$ and pseudo
wave functions $\ket{\bar{u}_{n\kvec}}$ with eigenvalues
$\bar{E}_{n\kvec}$. This ``bare'' term is the largest contribution to
the orbital magnetization and the other terms in
Eqs.~(\ref{equ:M_orb_pseudo_dbare})--(\ref{equ:M_orb_pseudo_ddia}) can
be considered corrections to the bare value. At first sight it might
appear that these correction terms are plagued from the same
ill-definedness  of the position operator $\rvec$ in periodic systems,
as described in Section~\ref{sec:problem}. But, the calculation of these
corrections is actually trivial---even in extended systems---since the
action of the non-local operators $V_\Rvec^{\text{NL}}$,
$F_\Rvec^{\text{NL}}$, and ${\bf E}_\Rvec^{\text{NL}}$ is non-zero only
inside spherical regions around the atom at $\Rvec$.

First calculations of the orbital magnetization for Fe, Co, and Ni have
been carried out in Ref.~[\citeonline{Ceresoli_2010a}] using this
pseudopotential approach. We report results for these calculations in
Section~\ref{sec:FeCoNi}, where Table~\ref{tab:FeCoNiresults} shows
explicit results for all terms that contribute to the orbital
magnetization in Eq.~(\ref{equ:M_orb_pseudo_tot}).

%%%%%%%%%%%%%%%%%%%%%%%%%%%%%%%%%%%%%%%%%%%%%%%%%%%%%%%%%%%%%%%%%%%%%%%%
\subsection{Single k-point derivation of the orbital magnetization}
\label{sec:k-point}

The expression for the orbital magnetization in
Eq.~(\ref{equ:m_orb_wannier_all}) is a $k$-space integral over the
Brillouin zone. The size of the Brillouin zone is inversely related to
the size of the unit cell in real space. As such, it is interesting to
study how large a supercell has to be such that the Brillouin-zone
integral is reasonably well approximated by a single $k$-point. Assuming
a large supercell, in Ref.~[\citeonline{Ceresoli_2007}] an effective
single $k$-point expression is derived, which is particularly useful
since most simulations for noncrystalline systems, including
Car-Parrinello simulations,\cite{Car_85} are routinely performed using a
large supercell in combination with only one $k$-point. We present a
simplified derivation here and refer the reader to the original
manuscript in  Ref.~[\citeonline{Ceresoli_2007}] for further details.

The starting point is Eq.~(\ref{equ:m_orb_wannier_all}), slightly
rewritten to include the antisymmetric tensor
$\epsilon_{\gamma\alpha\beta}$ and the shortcut
$\partial_\alpha\equiv\partial/\partial k_\alpha$,\footnote{The
corresponding expression, Eq.~(2) in Ref.~[\citeonline{Ceresoli_2007}],
has an erroneous minus sign, which propagates through the remainder of
that reference. Also, several sums over $n$ are missing in
Ref.~[\citeonline{Ceresoli_2007}].}
\begin{equation}
M_{\text{orb},\gamma} = \sum_{\alpha\beta, n}
\epsilon_{\gamma\alpha\beta}\;\frac{e}{2\hbar c}\,{\rm Im}\!
\int\frac{\dk}{(2\pi)^3}\;f_{n\kvec}\;
\me{\partial_\alpha u_{n\kvec}}{H_\kvec+E_{n\kvec}-2\mu}
{\partial_\beta u_{n\kvec}}\;.
\end{equation}
For a sufficiently large supercell of volume $V$, this expression can be
approximated by replacing the $k$-space integral by the single $k$-point
$\kvec=0$,
\begin{equation}\label{equ:single_k_start}
M_{\text{orb},\gamma} \approx \sum_{\alpha\beta, n}
\epsilon_{\gamma\alpha\beta}\;\frac{e}{2\hbar cV}
\,{\rm Im}\;f_{n0}\;\me{\partial_\alpha u_{n0}}
{H_0+E_{n0}-2\mu}{\partial_\beta u_{n0}}\;.
\end{equation}
This approximation is possible because the integrand for the orbital
magnetization itself is gauge invariant.\cite{Ceresoli_2006} Note that
the analog case of the electric polarization, where only the integral as
a whole is gauge invariant, is more complicated.\cite{Resta_96} If we
denote the shortest reciprocal lattice vectors of the supercell as
$\bvec_j$ and the derivative in the same directions as $\partial_j$, 
Eq.~(\ref{equ:single_k_start}) can be written as
\begin{equation}
\Mvec\orb \approx \sum_{ijl,n}
\epsilon_{ijl}\;\bvec_i|\bvec_j||\bvec_l|\;\frac{e}{2\hbar c(2\pi)^3}\,
{\rm Im}\;f_{n0}\;\me{\partial_j u_{n0}}{H_0+E_{n0}-2\mu}
{\partial_l u_{n0}}\;,
\end{equation}
where the $k$-space derivatives of $\ket{u_{n0}}$ can be understood as
the finite difference
\begin{equation}
\ket{\partial_l u_{n0}} = \lim\limits_{\lambda\to 0}
\frac{1}{\lambda|\bvec_l|}\big(\ket{u_{n\lambda\bvec_l}} -
\ket{u_{n0}}\big)\;.
\end{equation}
Such derivatives are numerically difficult due to gauge issues and
special care has to be taken as described in
Section~\ref{sec:k-space_derivatives}. Approximating this expression
with $\lambda=1$ and using the dual states defined in 
Eq.~(\ref{equ:dual_states}), the orbital magnetization becomes
\begin{equation}\label{equ:single_k_point_formula}
\Mvec\orb \approx \sum_{ijl,n}\epsilon_{ijl}\;\bvec_i\;
\frac{e}{2\hbar c(2\pi)^3}\,{\rm Im}\;f_{n0}\;
\me{\tilde{u}_{n\bvec_j}}{H_0+E_{n0}-2\mu}{\tilde{u}_{n\bvec_l}}\;.
\end{equation}
Note that the matrix elements $\me{u_{n0}}{\cdots}{u_{n0}}$ and
$\me{\tilde{u}_{n\bvec_j}}{\cdots}{u_{n0}}$, due to the orthogonality of
the dual states in Eq.~(\ref{equ:ortho_dual}), are purely real and
vanish after taking the imaginary part.

Since the periodicity in $k$-space is described by the vectors
$\bvec_j$, the next nearest neighbor points necessary for these
derivatives can all be related to the $\kvec=0$ point via
$\ket{u_{n\bvec_j}}=e^{-i\bvec_j\cdot\rvec}\ket{u_{n0}}$. The resulting
expression for the orbital magnetization is thus indeed a single
$k$-point formula in the sense that the Hamiltonian has to be
diagonalized only once at $\kvec=0$ to obtain the eigenvalues $E_{n0}$
and the states $\ket{u_{n0}}$; the orbital magnetization can then be
computed only based on these quantities.

\begin{figure}[t]
\begin{center}
\includegraphics[width=0.45\textwidth]{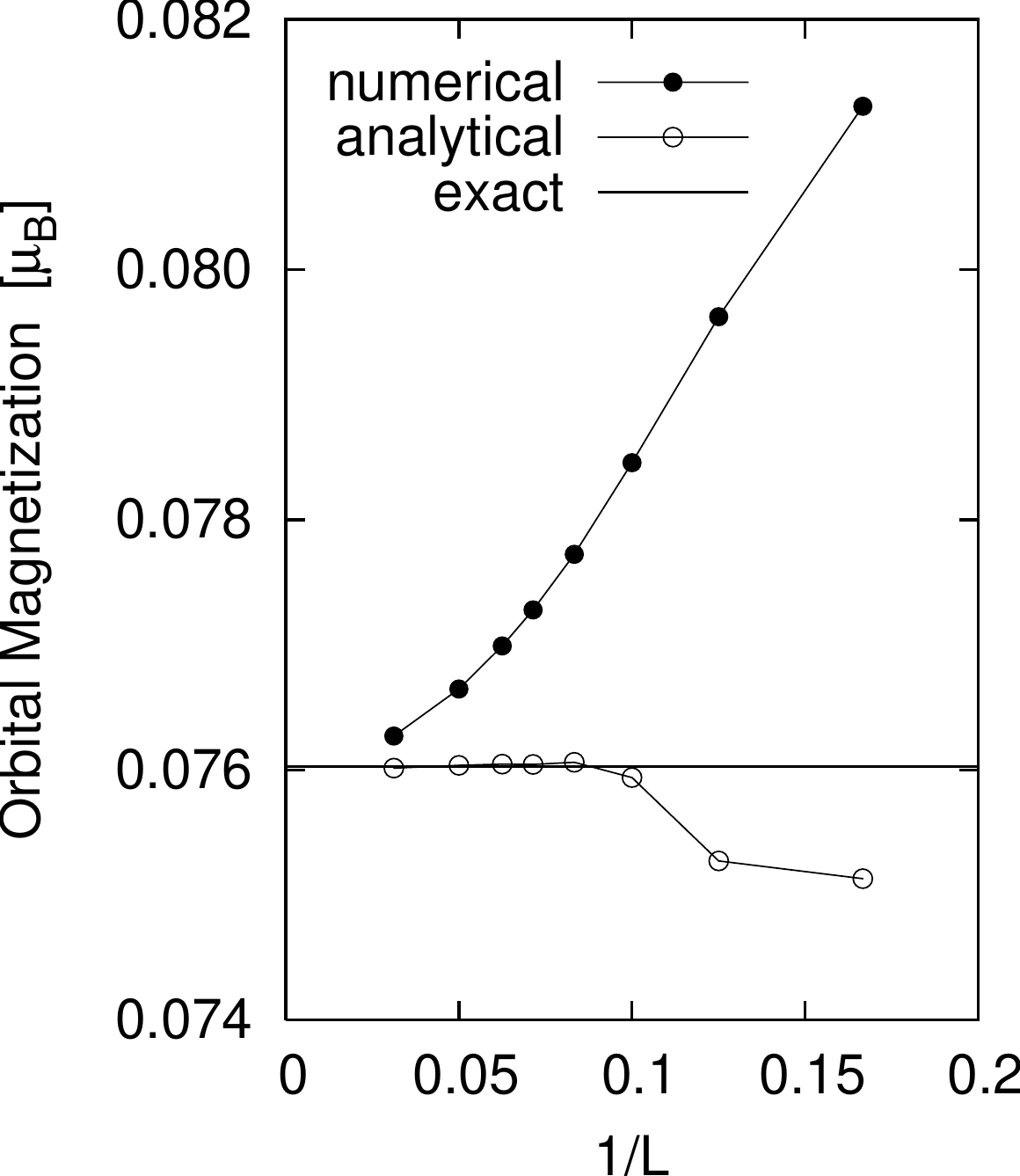}\\
\begin{minipage}{0.6\textwidth}
\caption{\label{fig:single_k_results}Orbital magnetization of the
Haldane model\cite{Haldane_88} in units of the Bohr magneton $\mu_B$ as
a function of the supercell size $L$. The exact value is compared with
the results obtained by evaluating the single $k$-point formula from
Eq.~(\ref{equ:single_k_point_formula}), using the \emph{numerical} and
\emph{analytical} methods for the $k$-space derivatives outlined in
Section~\ref{sec:k-space_derivatives}. (Reprinted with permission from
Ref.~[\citeonline{Ceresoli_2007}]; \copyright\ 2007 American Physical
Society).}
\end{minipage}
\end{center}
\end{figure}

The efficiency of this single $k$-point formula has been tested in the
context of a simple two-dimensional tight-binding model developed by
Haldane.\cite{Haldane_88} This model has no macroscopic magnetization
and breaks time-reversal symmetry through a complex second-nearest
neighbor hopping, resulting in staggered magnetic fluxes that cancel
throughout the unit cell. The convergence of the orbital magnetization
with respect to the size of the supercell can be seen in
Fig.~\ref{fig:single_k_results}.

%%%%%%%%%%%%%%%%%%%%%%%%%%%%%%%%%%%%%%%%%%%%%%%%%%%%%%%%%%%%%%%%%%%%%%%%
\subsection{Wannier interpolation of the orbital magnetization}
\label{sec:Wannier_interpolation}

An efficient, alternative numerical approach to evaluate the orbital
magnetization that targets metallic materials specifically is currently
being developed.\cite{Lopez_11} The basic idea is to first perform an
\emph{ab initio} ground-state calculation on a relatively coarse
$k$-mesh that yields ground-state wave functions and energies. From
these, partially occupied disentangled Wannier functions are
constructed,\cite{Souza01} which then serve as a compact
tight-binding-like parameterization of the \emph{ab initio} results. The
Wannier representation can then be used to interpolate quantities of
interest on (almost) arbitrarily fine $k$-meshes and evaluate the
corresponding Brillouin-zone integrals easily. This method of
\emph{Wannier interpolation} has been developed in the context of the
anomalous Hall effect in ferromagnets, where $k$-meshes of up to
$300\times 300\times 300$ grid points were necessary to converge the
Brillouin-zone integral for the anomalous Hall
conductivity\cite{Wang_06}---a quantity which has much in common with
the orbital magnetization. Although the Brillouin-zone integrand of the
orbital magnetization in Eqs.~(\ref{equ:m_orb_wannier_single}),
(\ref{equ:m_orb_wannier_all}), (\ref{equ:semi_classical}), and
(\ref{equ:M_orb_finite}) is not as spiky as the one for the anomalous
Hall conductivity, a Wannier interpolation scheme still provides
advantages.  At the time of writing, the corresponding manuscript
discussing the interpolation of the orbital magnetization in
Ref.~[\citeonline{Lopez_11}] is under review, and we refer the reader to
the published version---once available---for discussion, results, and
further details.

%%%%%%%%%%%%%%%%%%%%%%%%%%%%%%%%%%%%%%%%%%%%%%%%%%%%%%%%%%%%%%%%%%%%%%%%
\subsection{DFT specific aspects of calculating the orbital
magnetization}
\label{sec:DFT}

The formalisms collected in this review article describe the calculation
of the orbital magnetization in periodic solids. As such, they can be
used in conjunction with any first-principles method suitable for
solids. While other successful approaches exist, density functional
theory (DFT)\cite{Kohn_64} has been the primary tool for modeling many
molecules and solids~\cite{Richard_04}; in fact, the vast majority of
all electronic-structure calculations today are DFT \cite{80}. It is
thus appropriate to consider some aspects of the calculation of the
orbital magnetization that are DFT specific.

One of the most important ``ingredients'' for DFT calculations is the
exchange-correlation functional used. As such, it is interesting to
study which exchange-correlation functional gives results for the
orbital magnetization in best agreement with experiment. However, since
the theory of orbital magnetization in solids is  rather young, a
systematic study of the orbital magnetization's dependence on the
functional has not been performed yet. The results for Fe, Co, and Ni
discussed in the next section suggest a slightly better performance of
PBE\cite{PBE} compared to LDA,\cite{PZ81} but that might also be related
to PBE's better overall performance in solids.

However, even if we had access to the \emph{true} exchange-correlation
functional, DFT is only guaranteed to produce the correct ground-state
density; a similar claim for the orbital currents cannot be made. Thus,
in principle, the calculation of the orbital magnetization---which is
defined in terms of orbital currents---within DFT is not guaranteed to
yield correct results. It seems that a formulation that also includes
currents, such as current density functional theory
(CDFT),\cite{Vignale_88} should systematically improve results for the
orbital magnetization. But, it
remains to be seen if appropriate current functionals, in conjunction
with CDFT, do indeed yield improved results over standard DFT.
Alternatively, time-dependent current density functional theory
(TD-CDFT)\cite{Ghosh_88, Vignale_04} or the better developed
time-dependent density functional theory (TD-DFT)\cite{Runge_84} could
in principle be employed. But again, at this point it is not clear that
they provide any practical advantage for calculating the orbital
magnetization over standard DFT. A somewhat more detailed discussion of
these aspects is given in Ref.~[\citeonline{Ceresoli_2006}].

%%%%%%%%%%%%%%%%%%%%%%%%%%%%%%%%%%%%%%%%%%%%%%%%%%%%%%%%%%%%%%%%%%%%%%%%
\section{Orbital Magnetization in Fe, Co, and Ni}
\label{sec:FeCoNi}
%%%%%%%%%%%%%%%%%%%%%%%%%%%%%%%%%%%%%%%%%%%%%%%%%%%%%%%%%%%%%%%%%%%%%%%%

Calculations of the orbital magnetization using the full theory have
first been reported in Ref.~[\citeonline{Ceresoli_2010a}]. In that
paper, the authors calculate the orbital magnetization of Fe, Co, and Ni
within DFT by evaluating Eq.~(\ref{equ:m_orb_wannier_all}) in the
pseudopotential context of Section~\ref{sec:pseudopotentials}, using the
linear-response method for the $k$-space derivative described in
Section~\ref{sec:k-space_derivatives}. This framework was implemented in
the first-principles electronic-structure code \textsc{PWscf}, which is
part of the \textsc{Quantum-Espresso} package.\cite{QE} At the time of
writing, this implementation is not yet publicly available.

\begin{table}[t]
\tbl{\label{tab:FeCoNiresults} Orbital magnetization of Fe, Co, and Ni
in units of the Bohr magneton $\mu_B$ per atom parallel to the spin, for
different spin directions and different exchange-correlation (XC)
functionals. The total orbital magnetization is given by $\Mvec\orb =
\Mvec_{\text{orb}}^{\text{bare}} +
\Delta\Mvec_{\text{orb}}^{\text{bare}} +
\Delta\Mvec_{\text{orb}}^{\text{para}} +
\Delta\Mvec_{\text{orb}}^{\text{dia}}$, where the contributions are
defined in
Eqs.~(\ref{equ:M_orb_pseudo_tot})--(\ref{equ:M_orb_pseudo_ddia}). Values
are taken from Ref.~[\citeonline{Ceresoli_2010a}] and its supplemental
materials; see footnote in main text. Experimental values are taken from
Refs.~[\citeonline{Meyer_61}] and [\citeonline{Wu_01}]. (Reprinted with
permission from Ref.~[\citeonline{Ceresoli_2010a}]; \copyright\ 2010
American Physical Society).}
{\begin{tabular*}{\textwidth}{@{\extracolsep{\fill}}llccccccr@{}}\Hline\\[-1.8ex]
Metal & \multicolumn{1}{c}{dir.} & exp. & XC & $M\orb^{}$ & $M_{\text{orb}}^{\text{bare}}$ & $\Delta M_{\text{orb}}^{\text{bare}}$
& $\Delta M_{\text{orb}}^{\text{para}}$ & $\Delta M_{\text{orb}}^{\text{dia}}$\\[0.8ex]\hline\\[-1.8ex]
\mr{4}{{\it bcc}-Fe} & \mr{2}{[001]$^*$} & \mr{2}{0.081} & PBE & 0.0658 & 0.0639 & 0.0000 & 0.0016 & 0.0003\\
                     &                   &               & LDA & 0.0642 & 0.0616 & 0.0005 & 0.0017 & 0.0004\\\cline{2-9}
                     & \mr{2}{[111]}     & \mr{2}{---}   & PBE & 0.0660 & 0.0637 & 0.0005 & 0.0015 & 0.0003\\
                     &                   &               & LDA & 0.0633 & 0.0609 & 0.0005 & 0.0015 & 0.0004\\\hline
\mr{4}{{\it fcc}-Co} & \mr{2}{[111]$^*$} & \mr{2}{0.120} & PBE & 0.0756 & 0.0722 & 0.0018 & 0.0014 & 0.0002\\
                     &                   &               & LDA & 0.0741 & 0.0706 & 0.0019 & 0.0014 & 0.0002\\\cline{2-9}
                     & \mr{2}{[001]}     & \mr{2}{---}   & PBE & 0.0660 & 0.0629 & 0.0016 & 0.0013 & 0.0002\\
                     &                   &               & LDA & 0.0642 & 0.0608 & 0.0016 & 0.0016 & 0.0002\\\hline
\mr{4}{{\it hcp}-Co} & \mr{2}{[001]$^*$} & \mr{2}{0.133} & PBE & 0.0957 & 0.0908 & 0.0032 & 0.0014 & 0.0003\\
                     &                   &               & LDA & 0.0924 & 0.0875 & 0.0032 & 0.0014 & 0.0003\\\cline{2-9}
                     & \mr{2}{[100]}     & \mr{2}{---}   & PBE & 0.0867 & 0.0822 & 0.0028 & 0.0015 & 0.0003\\
                     &                   &               & LDA & 0.0837 & 0.0792 & 0.0029 & 0.0013 & 0.0003\\\hline
\mr{4}{{\it fcc}-Ni} & \mr{2}{[111]$^*$} & \mr{2}{0.053} & PBE & 0.0519 & 0.0494 & 0.0017 & 0.0007 & 0.0001\\
                     &                   &               & LDA & 0.0545 & 0.0519 & 0.0019 & 0.0007 & 0.0000\\\cline{2-9}
                     & \mr{2}{[001]}     & \mr{2}{---}   & PBE & 0.0556 & 0.0527 & 0.0022 & 0.0006 & 0.0001\\
                     &                   &               & LDA & 0.0533 & 0.0505 & 0.0020 & 0.0007 & 0.0001\\[0.8ex]\Hline\\[-1.8ex]
\multicolumn{9}{@{}l}{$^*$ denotes the experimental easy axis.}\\
\end{tabular*}}
\end{table}

Results for the orbital magnetization and its split-up into the bare and
correction terms according to
Eqs.~(\ref{equ:M_orb_pseudo_tot})--(\ref{equ:M_orb_pseudo_ddia}) are
collected in Table~\ref{tab:FeCoNiresults}; values for $M\orb$ are taken
from Ref.~[\citeonline{Ceresoli_2010a}] and values for the bare and
correction terms are taken from the supplemental materials of the same
reference.\footnote{Reference~[\citeonline{Ceresoli_2010a}] and its
supplemental materials only contain PBE results. The values for LDA
results were obtained from the authors of
Ref.~[\citeonline{Ceresoli_2010a}] through personal communication.} The
results clearly show that, as pointed out in
Section~\ref{sec:importance}, the orbital magnetization in Fe, Co, and
Ni is indeed a small effect of only a few percent compared to their spin
magnetization of 2.083~$\mu_B$ (Fe), 1.523~$\mu_B$ (Co), and 
0.518~$\mu_B$ (Ni).\cite{LB} It is also interesting to see that the
orbital magnetization is well described by the bare term 
$M_{\text{orb}}^{\text{bare}}$ alone and the pseudopotential corrections
$\Delta M_{\text{orb}}^{\text{bare}}$, $\Delta
M_{\text{orb}}^{\text{para}}$, and $\Delta M_{\text{orb}}^{\text{dia}}$
are usually small. Although only a small effect, it can also be seen
from these results that PBE gives slightly better results than LDA.
However, that might also be related to the better performance of PBE in
solids in general.

With the full theory of orbital magnetization available, we can now
assess the accuracy of the often-used muffin-tin approximation,
mentioned in Section~\ref{sec:problem}. In this approximation, the
orbital contribution can be simply calculated using
Eq.~(\ref{equ:magnetization_finite}), integrating over non-overlapping
spheres centered around the atoms. Since the sphere volumes are finite
systems, the problem of the ill-defined position operator is
circumvented. In particular, one can calculate
\begin{eqnarray}\label{equ:MT_approx}
\Mvec\orb^{\text{MT}} &=& -\frac{e}{2cV_0}\sum_n
                          \int\frac{\dk}{(2\pi)^3} \,f_{n\kvec}\,
                          \me{u_{n\kvec}}{\rvec\times\vvec}
                          {u_{n\kvec}}_\text{\raisebox{-1.25ex}{MT}}\\
                      &=& -\frac{e}{2cV_0}\sum_n\int\frac{\dk}{(2\pi)^3}
                          \int_{\text{MT}}\!\!\!\dr\;f_{n\kvec}^{}\;
                          u_{n\kvec}^*(\rvec)\;\big[
                          \rvec\times(-i\hbar\nabla+\hbar\kvec)/m_e
                          \big]\;u_{n\kvec}(\rvec)\;,\nonumber
\end{eqnarray}
where we integrate inside all muffin-tin (MT) spheres in real space.
This muffin-tin contribution was calculated in
Ref.~[\citeonline{Ceresoli_2010a}] for a sphere radius of 2 Bohr. The
interstitial contribution is then defined as the total orbital
magnetization minus the muffin-tin contribution. Results for Fe, Co, and
Ni are collected in Table~\ref{tab:FeCoNi_int_MT}. Surprisingly, the
results show that up to 34\% of the orbital magnetization in Fe, up to
16\% in Co, and still up to 8\% in Ni originate from the interstitial.
As such, it is not surprising that calculations of the orbital
magnetization---using the muffin-tin approximation---strongly
underestimate the effect.\cite{Wu_01, Sharma_07}. These results, \emph{a
posteriori}, provide indeed a strong justification for developing the
complete theory of orbital magnetization.

\begin{table}[t]
\tbl{\label{tab:FeCoNi_int_MT} Orbital magnetization of Fe, Co, and Ni
in $\mu_B$/atom parallel to the spin, for different spin directions. The
total orbital magnetization is here split up into contributions coming
from the muffin-tin spheres, i.e.\ Eq.~(\ref{equ:MT_approx}), and the
interstitial. Calculated values are taken from
Ref.~[\citeonline{Ceresoli_2010a}] and experimental values are from
Refs.~[\citeonline{Meyer_61}] and [\citeonline{Wu_01}]. (Reprinted with
permission from Ref.~[\citeonline{Ceresoli_2010a}]; \copyright\ 2010
American Physical Society).}
{\begin{tabular*}{0.75\textwidth}{@{\extracolsep{\fill}}llcccr@{}}
\Hline\\[-1.8ex]
Metal & \multicolumn{1}{c}{dir.} & exp. & $M\orb^{}$ & interstitial &
muffin tin\\[0.8ex]\hline\\[-1.8ex]
\mr{2}{{\it bcc}-Fe} & [001]$^*$ & 0.081 & 0.0658 & 0.0225 & 0.0433\\
                     & [111]     & ---   & 0.0660 & 0.0216 & 0.0444\\
\mr{2}{{\it fcc}-Co} & [111]$^*$ & 0.120 & 0.0756 & 0.0122 & 0.0634\\
                     & [001]     & ---   & 0.0660 & 0.0064 & 0.0596\\
\mr{2}{{\it hcp}-Co} & [001]$^*$ & 0.133 & 0.0957 & 0.0089 & 0.0868\\
                     & [100]     & ---   & 0.0867 & 0.0068 & 0.0799\\
\mr{2}{{\it fcc}-Ni} & [111]$^*$ & 0.053 & 0.0519 & 0.0008 & 0.0511\\
                     & [001]     & ---   & 0.0556 & 0.0047 & 0.0509
		       \\[0.8ex]\Hline\\[-1.8ex]
\multicolumn{6}{@{}l}{$^*$ denotes the experimental easy axis.}\\
\end{tabular*}}
\end{table}

The overall results for the orbital magnetization in
Table~\ref{tab:FeCoNi_int_MT} have also been confirmed with an
independent implementation of the Wannier interpolation method presented
in Section~\ref{sec:Wannier_interpolation} and 
Ref.~[\citeonline{Lopez_11}]. On the other hand, preliminary
calculations by another group, using an implementation in the
\textsc{Wien2k} code,\cite{Wien2k} disagree with the results in
Table~\ref{tab:FeCoNi_int_MT}; however, those results were never
published.\cite{Yao_2009}

%%%%%%%%%%%%%%%%%%%%%%%%%%%%%%%%%%%%%%%%%%%%%%%%%%%%%%%%%%%%%%%%%%%%%%%%
\section{Applications of the Orbital Magnetization}
\label{sec:applications}
%%%%%%%%%%%%%%%%%%%%%%%%%%%%%%%%%%%%%%%%%%%%%%%%%%%%%%%%%%%%%%%%%%%%%%%%

While the orbital magnetization of materials itself is a useful
concept, its \emph{change} is of much practical interest, as it can be
related to several experimental probes.  The calculation of changes of
the orbital magnetization has been possible for about 15 years now,
using linear-response methods.\cite{Mauri_96a, Mauri_96b, Pickard_02,
Sebastiani_01, Sebastiani_02} But, now that the orbital magnetization
itself can be calculated, numerous possibilities  open up to develop new
approaches---much simpler than linear-response methods---using finite
differences. Such approaches might prove particularly useful in
situations where linear-response calculations are cumbersome or
impossible, or in conjunction with more complex methods of treating
exchange and correlation effects, such as hybrid functionals, DFT+U,
exact exchange, or beyond-DFT approaches.  In the following sections we
focus on calculating the nuclear magnetic resonance shielding tensor and
the electron paramagnetic resonance $g$-tensor as a finite difference of
the orbital magnetization.

%%%%%%%%%%%%%%%%%%%%%%%%%%%%%%%%%%%%%%%%%%%%%%%%%%%%%%%%%%%%%%%%%%%%%%%%
\subsection{Nuclear magnetic resonance as a derivative\\
of the orbital magnetization}
\label{sec:NMR}

Nuclear magnetic resonance (NMR) is one of the most important
experimental techniques used to determine the structure of molecules,
liquids, and other disordered systems; it has thus evolved into one of
the most widely used methods in structural chemistry.\cite{Rabi,
NMR_encyclopedia} Similar to the orbital magnetization itself, methods
to calculate the NMR response for molecules and clusters, i.e.\ finite
systems, were developed early on in the quantum-chemistry
community,\cite{Kutzelnigg_90} but calculations for periodic solids were
impossible, as the inclusion of a constant magnetic field requires a
vector potential that breaks translational symmetry. While
linear-response frameworks to calculate the NMR shielding in solids have
been available for a while,\cite{Mauri_96b, Sebastiani_01,
Sebastiani_02} knowledge of the orbital magnetization itself provides a
much simpler way to calculate the shielding, referred to as the
\emph{converse} NMR approach. The corresponding formalism is described
in detail in  Ref.~[\citeonline{Thonhauser_2009a}].

When a sample is put in an external magnetic field $\Bvec^{\text{ext}}$,
a field $\Bvec^{\text{ind}}$ is induced, resulting in a total magnetic
field of $\Bvec_i^{\text{tot}}=\Bvec^{\text{ind}}_i+\Bvec^{\text{ext}}$
at the site of atom $i$. The NMR shielding tensor
$\boldsymbol{\sigma}_i$ for atom $i$ is then defined as
$\Bvec_i^{\text{ind}}=-\boldsymbol{\sigma}_i\cdot\Bvec^{\text{ext}}$, or
equivalently as
\begin{equation}\label{equ:sigma_def}
\sigma_{\alpha\beta,i} = -\frac{\partial B_{\alpha,i}^{\text{ind}}}
{\partial B_{\beta}^{\text{ext}}}\;.
\end{equation}
We now consider an artificial magnetic dipole $\mvec_i$ at site $i$.
Such a dipole would have the energy
$E=-\mvec_i^{}\cdot\Bvec_i^{\text{tot}}$, which leads to
\begin{equation}\label{equ:B_tot}
B_{\alpha,i}^{\text{tot}} = -\partial E/\partial m_{\alpha,i}\;.
\end{equation}
Starting from Eq.~(\ref{equ:sigma_def}), we can now use a thermodynamic
relationship between mixed partial derivatives and insert
Eq.~(\ref{equ:B_tot}) to find
\begin{eqnarray}\label{equ:sigma_M_orb}
\sigma_{\alpha\beta,i}
   &=& -\frac{\partial\big(B_{\alpha,i}^{\text{tot}}-
       B_{\alpha}^{\text{ext}})}{\partial B_{\beta}^{\text{ext}}} =
       \delta_{\alpha\beta}+\frac{\partial}{\partial
       B_\beta^{\text{ext}}}\frac{\partial E}
       {\partial m_{\alpha,i}}\nonumber\\
   &=& \delta_{\alpha\beta}+\frac{\partial}{\partial m_{\alpha,i}}
       \frac{\partial E}{\partial B_\beta^{\text{ext}}} =
       \delta_{\alpha\beta}-V_0\frac{\partial M_{\text{orb},\beta}}
       {\partial m_{\alpha,i}}\;,
\end{eqnarray}
where $V_0$ is the unit-cell volume and we have written the macroscopic
orbital magnetization as $ M_{\text{orb},\beta}=-(1/V_0)\,\partial
E/\partial B_\beta^{\text{tot}}$. By doing so, in the last step we have
made the assumption that the external field can be replaced by the total
field, which corresponds to a particular choice of geometry for the
sample shape. However, the results for other sample shapes can easily be
recovered through knowledge of the
susceptibility.\cite{Thonhauser_2009a}

Equation~(\ref{equ:sigma_M_orb}) expresses the shielding tensor in terms
of changes in the orbital magnetization due to the presence of the
artificial dipole $\mvec_i$, and finite differences can readily be
calculated. To that end, the dipole $\mvec_i$ needs to be included in
the system's Hamiltonian via its vector potential, for which a suitable
form has been derived\cite{Thonhauser_2009a}---a minimal change to
existing codes, compared to a linear-response framework. Note that this
approach  circumvents the difficulties related to including finite
external magnetic fields and the choice of gauge origin.   In order to
evaluate Eq.~(\ref{equ:sigma_M_orb}) in a pseudopotential framework, a
GIPAW transformation similar to Section~\ref{sec:pseudopotentials} has
to be performed, which includes the vector potential of the artificial
dipole in the Hamiltonian; the corresponding framework has been derived
in Ref.~[\citeonline{Ceresoli_2010b}].

The converse NMR approach has been implemented in the first-principles
electronic-structure code \textsc{PWscf}, which is part of the
\textsc{Quantum-Espresso} package.\cite{QE} At the time of writing, this
implementation is not yet publicly available. Calculations have been
performed within DFT by evaluating Eq.~(\ref{equ:sigma_M_orb}) through
finite differences in the GIPAW pseudopotential context  using the
linear-response method for the $k$-space derivative described in
Section~\ref{sec:k-space_derivatives}. This implementation has already
been successfully applied to calculate H, C, F, O, P, and Si NMR shifts
in a variety of systems such as small molecules, polycyclic aromatic
hydrocarbons, bulk water, and selected SiO$_2$
crystals.\cite{Thonhauser_2009a , Ceresoli_2010b, Thonhauser_2009b}

%%%%%%%%%%%%%%%%%%%%%%%%%%%%%%%%%%%%%%%%%%%%%%%%%%%%%%%%%%%%%%%%%%%%%%%%
\subsection{Electron paramagnetic resonance as a\\
change of the orbital magnetization}
\label{sec:EPR}

Electron paramagnetic resonance (EPR) is in nature similar to NMR,
except that the focus is on electronic spins instead of nucleic spins.
Most stable molecules have no unpaired electrons, such that EPR is
limited to paramagnetic materials and is thus less widely used. The
property of interest is then the electron $g$-tensor. As in the case of
NMR, a linear-response framework exists to calculate the
$g$-tensor.\cite{Pickard_02} However, knowledge of the orbital
magnetization allows for a direct calculation.\cite{Ceresoli_2010a}

The $g$-tensor deviation $\Delta g_{\alpha\beta}$ from the free-electron
value of $g_e=2.002\,319$ can be calculated through a change in the
orbital magnetization as\cite{Ceresoli_2010a, Schreckenbach_97}
\begin{equation}\label{equ:EPR}
\Delta g_{\alpha\beta} = -\frac{1}{\mu_B}\,\frac{\partial}
{\partial S_\beta}M_{\text{orb},\alpha}\;,
\end{equation}
where $S$ is the total spin. This approach has been used to study the
microscopic structure of radicals and paramagnetic defects in
solids.\cite{Ceresoli_2010a}. In particular, calculations have been
performed at the DFT level by evaluating Eq.~(\ref{equ:EPR}) through
finite differences according to a spin flip in the GIPAW pseudopotential
context described in Section~\ref{sec:pseudopotentials}  using the
linear-response method for the $k$-space derivative described in
Section~\ref{sec:k-space_derivatives}. It is interesting to note that,
while for the orbital magnetization the pseudopotential correction terms
in Eqs.~(\ref{equ:M_orb_pseudo_dbare})--(\ref{equ:M_orb_pseudo_ddia})
play only a minor role (see Table~\ref{tab:FeCoNiresults} on page
\pageref{tab:FeCoNiresults}), the correction terms play a more
significant role when calculating $\Delta g_{\alpha\beta}$.

%%%%%%%%%%%%%%%%%%%%%%%%%%%%%%%%%%%%%%%%%%%%%%%%%%%%%%%%%%%%%%%%%%%%%%%%
\subsection{Other derivatives of the orbital magnetization}
\label{sec:other_derivatives}

Derivatives of the orbital magnetization also play an important role in
other areas of condensed matter physics. We conclude this section with a
non-exhaustive list, mentioning several recent examples.

The magnetic susceptibility of a material is defined as the derivative
of the total magnetization with respect to an external magnetic field.
As such, the derivative of the orbital magnetization contributes to the
magnetic susceptibility in general.\cite{Mauri_96a}

Furthermore, in Ref.~[\citeonline{Murakami_2006}] the author links the
spin-Hall conductivity in insulators through a St\v{r}eda-like formula
to the magnetic susceptibility, and in particular, to the derivative of
the orbital magnetization with respect to a magnetic field. Finite
differences calculations of the orbital magnetization were carried out
for a simple two-band model of graphene evaluating
Eq.~(\ref{equ:m_orb_wannier_all}).

As pointed out in Ref.~[\citeonline{Cooper_2009}], it might be possible
to identify non-abelian quantum Hall states by experimental measurements
of the temperature dependence of the orbital magnetization. This is of
particular interest in view of the recently developed analytical formula
for the temperature dependent orbital magnetization in
Eq.~(\ref{equ:M_orb_finite}).

In insulators with broken time-reversal and inversion symmetry, in first
order an electric field $\Evec$ can induce a magnetization $\Mvec$ and a
magnetic field $\Bvec$ can induce an electric polarization $\Pvec$. The
coupling is described by the linear magnetoelectric polarizability
$a_{\alpha\beta}$,\cite{Essin_2009} defined as
\begin{equation}
a_{\alpha\beta} = \frac{\partial M_\beta}{\partial E_\alpha}
\bigg|_{\Bvec=0} = \frac{\partial P_\alpha}{\partial B_\beta}
\bigg|_{\Evec=0}\;.
\end{equation}
Orbital contributions to these derivatives are discussed in
Refs.~[\citeonline{Malashevich_2010}] and [\citeonline{Essin_2010}].

%%%%%%%%%%%%%%%%%%%%%%%%%%%%%%%%%%%%%%%%%%%%%%%%%%%%%%%%%%%%%%%%%%%%%%%%
\section{Conclusions and Outlook}
\label{sec:Conclusions}
%%%%%%%%%%%%%%%%%%%%%%%%%%%%%%%%%%%%%%%%%%%%%%%%%%%%%%%%%%%%%%%%%%%%%%%%

The theory of orbital magnetization in solids is still in its infancy,
being developed only in 2005. Its ``older brother''---the modern theory
of electric polarization---has been around for over a decade
longer.\cite{King-Smith_93, Vanderbilt_93} The modern theory of electric
polarization was hugely successful, as the number of citations of its
fundamental papers shows. The theory was a breakthrough because it
allows calculations of the polarization as a bulk property, without the
need for expensive calculations on slabs or clusters. It is difficult to
predict whether the theory of orbital magnetization will enjoy the same
level of success and widespread use; in light of the fact that magnetism
in many materials is well described with the spin contribution only, for
which theories have existed for a long time, it seems unlikely that it
will.

However, the true success of the modern theory of electric polarization
came about when its formalism was widely available in first-principles
electronic-structure codes. In fact, today its formalism is present in
almost all of the popular and widely used codes.\cite{Resta_2010}
Currently, the situation is different for the theory of orbital
magnetization. At the time of writing, prototype implementations only
exist for the \textsc{PWscf} part of the \textsc{Quantum-Espresso}
package\cite{QE} and \textsc{Wien2k}\cite{Wien2k}, which are not even
publicly available yet. Implementations in \textsc{Vasp},\cite{VASP}
\textsc{Abinit},\cite{ABINIT} and \textsc{Adf}\cite{ADF} are in a
development stage. Once more implementations become widely available,
the theory of orbital magnetization can be expected to play an important
role in condensed matter physics and  materials science.

%%%%%%%%%%%%%%%%%%%%%%%%%%%%%%%%%%%%%%%%%%%%%%%%%%%%%%%%%%%%%%%%%%%%%%%%
\section*{Acknowledgements}
%%%%%%%%%%%%%%%%%%%%%%%%%%%%%%%%%%%%%%%%%%%%%%%%%%%%%%%%%%%%%%%%%%%%%%%%

This review article was written while on sabbatical at the Department of
Materials, University of Oxford. I would like to thank Wake Forest
University for providing this opportunity and \emph{Prof.\ Nicola
Marzari} for being a wonderful host during that time. I am truly
indebted to \emph{Dr.\ Davide Ceresoli} for countless discussions
concerning the orbital magnetization, for proofreading this manuscript, 
and for his everlasting patience with my questions. I am further
grateful for enlightening discussions with Profs. \emph{David
Vanderbilt, Raffaele Resta}, and \emph{Qian Niu}. Finally, I am also
grateful to \emph{Loah Stevens}, who helped with the literature search
and organization of the references for this review article.

%%%%%%%%%%%%%%%%%%%%%%%%%%%%%%%%%%%%%%%%%%%%%%%%%%%%%%%%%%%%%%%%%%%%%%%%
\section*{References}
%%%%%%%%%%%%%%%%%%%%%%%%%%%%%%%%%%%%%%%%%%%%%%%%%%%%%%%%%%%%%%%%%%%%%%%%


\begin{thebibliography}{999}

\bibitem{history_1} Daniel C. Mattis, \emph{The Theory of Magnetism
   Made Simple: An Introduction to Physical Concepts and to Some Useful
   Mathematical Methods}, p.\ 1 (World Scientific Publishing, Singapore,
   2006). The first chapter of this book gives a very nice introduction
   to the history of magnetism and is available free of charge through
   the Wold Scientific website at
   \url{http://www.worldscibooks.com/etextbook/5372/5372\_chap01.pdf}.

\bibitem{history_2} Lucretius Carus, \emph{De Rerum Natura}, 1st
   century B.C. References are to vv. 906 ff., in the translation by Th.
   Creech, London, 1714. Pliny, quoted in W. Gilbert, \emph{De Magnete},
   trans. (Gilbert Club, London, 1900), rev. ed., p.\ 8 (Basic Books,
   New York, 1958).    

\bibitem{Meyer_61} A.\,J.\,P. Meyer and G. Asch, J. Appl. Phys.
   {\bf 32}, S330 (1961).

%80
\bibitem{Qiao_2004} S. Qiao, A. Kimura, H. Adachi, K. Iori, K.
   Miyamoto, T. Xie, H. Namatame, M. Taniguchi, A. Tanaka, T. Muro,
   S. Imada, and S. Suga, Phys. Rev. B {\bf 70}, 134418 (2004).

%89
\bibitem{Gotsis_2003} H.\,J. Gotsis and I.\,I. Mazin, Phys. Rev. B
   {\bf 68}, 224427 (2003).

%97
\bibitem{Taylor_2002} J.\,W. Taylor, J.\,A. Duffy, A.\,M. Bebb, M.\,R.
   Lees, L. Bouchenoire, S.\,D. Brown, and M.\,J. Cooper, Phys. Rev. B
   {\bf 66}, 161319(R) (2002).

%15
\bibitem{Thonhauser_2009a} T. Thonhauser, D. Ceresoli, A.\,A. Mostofi,
   N. Marzari, R. Resta, and D. Vanderbilt, J. Chem. Phys. {\bf 131},
   101101 (2009).

%9
\bibitem{Ceresoli_2010a} D. Ceresoli, U. Gerstmann, A.\,P. Seitsonen,
   and F. Mauri, Phys. Rev. B {\bf 81}, 060409(R) (2010).

%4
\bibitem{Malashevich_2010} A. Malashevich, I. Souza, S. Coh, and D.
   Vanderbilt, New J. Phys. {\bf 12}, 053032 (2010).

%6
\bibitem{Essin_2010} A.\,M. Essin, A.\,M. Turner, J.\,E. Moore, and
   D. Vanderbilt, Phys. Rev. B {\bf 81}, 205104 (2010).

%17
\bibitem{Essin_2009} A.\,M. Essin, J.\,E. Moore, and D.
   Vanderbilt, Phys. Rev. Lett. {\bf 102}, 146805 (2009).

%30
\bibitem{Murakami_2006} S. Murakami, Phys. Rev. Lett. {\bf 97},
   236805 (2006).

%41
\bibitem{Cooper_2009} N.\,R. Cooper and A. Stern, Phys. Rev. Lett.
   {\bf 102}, 176807 (2009).

\bibitem{King-Smith_93} R.\,D. King-Smith and D. Vanderbilt, Phys.
   Rev. B {\bf 47}, R1651 (1993).
   
\bibitem{Vanderbilt_93} D. Vanderbilt and R.\,D. King-Smith, Phys.
   Rev. B {\bf 48}, 4442 (1993).

\bibitem{Berry_84} M.\,V. Berry, Proc. R. Soc. London, Ser.
   A {\bf 392}, 45 (1984). 

%3
\bibitem{Xiao_2010} D. Xiao, M.-C. Chang, and Q. Niu, Rev. Mod. Phys.
   {\bf 82}, 1959 (2010).

%8
\bibitem{Resta_2010} R. Resta, J. Phys.: Condens. Matt. {\bf 22},
   123201 (2010).

%38
\bibitem{Blonski_2010} P. Blo\'nski, A. Lehnert, S. Dennler,
   S. Rusponi, M. Etzkorn, G. Moulas, P. Bencok, P. Gambardella,
   H. Brune, and J. Hafner, Phys. Rev. B {\bf 81}, 104426 (2010).

%104
\bibitem{Lazarovits_2002} B. Lazarovits, L. Szunyogh, and P.
   Weinberger, J. Magn. Magn. Matt. {\bf 240}, 331 (2002).

%96
\bibitem{Aldea_2003} A. Aldea, V. Moldoveanu, M. Nita, A. Manolescu,
   V. Gudmundsson, and B. Tanatar, Phys. Rev. B {\bf 67}, 035324 (2003).

%129
\bibitem{Durr_1999} H.\,A. D\"urr, S.\,S. Dhesi, E. Dudzik, D. Knabben,
   G. van der Laan, J.\,B. Goedkoop, and F.\,U. Hillebrecht, Phys. Rev.
   B {\bf 59}, R701 (1999).

%49
\bibitem{Bartolome_2008} J. Bartolom\'e, M. Garc\'ia, F. Bartolom\'e,
   F. Luis, R. L\'opez-Ruiz, F. Petroff, C. Deranlot, F. Wilhelm, A.
   Rogalev, P. Bencok, N.\,B. Brookes, L. Ruiz, and J.\,M.
   Gonz\'alez-Calbet, Phys. Rev. B {\bf 77}, 184420 (2008).

%90
\bibitem{Gambardella_2003} P. Gambardella, J. Phys.: Condens. Matt.
   {\bf 15}, S2533 (2003).

%82
\bibitem{Hong_2004} J. Hong and R.\,Q. Wu, Phys. Rev. B {\bf 70},
   060406(R) (2004).

%85
\bibitem{Lazarovits_2004} B. Lazarovits, L. Szunyogh, and P.
   Weinberger, J. Magn. Magn. Matt. {\bf 272--276}, 1658 (2004).

%102
\bibitem{Gambardella_2002} P. Gambardella, A. Dallmeyer, K. Maiti,
   M.\,C. Malagoli, W. Eberhardt, K. Kern, and C. Carbone, Nature
   {\bf 416}, 301 (2002).

%18
\bibitem{Oka_2009} T. Oka and H. Aoki, Phys. Rev. B {\bf 79}, 081406(R)
   (2009).

%50
\bibitem{Liu_2008} J. Liu, Z. Ma, A.\,R. Wright, and C. Zhang,
   J. Appl. Phys. {\bf 103}, 103711 (2008).

%23
\bibitem{Xiao_2007} D. Xiao, W. Yao, and Q. Niu, Phys. Rev. Lett.
   {\bf 99}, 236809 (2007).

%22
\bibitem{Yao_2008} W. Yao, D. Xiao, and Q. Niu, Phys. Rev. B {\bf 77},
   235406 (2008).

%71
\bibitem{Faulhaber_2005} D.\,R. Faulhaber and H.\,W. Jiang, Phys.
   Rev. B {\bf 72}, 233308 (2005).

%117
\bibitem{Oppen_2000} F. von Oppen, D. Ullmo, and H.\,U. Baranger,
   Phys. Rev. B {\bf 62}, 1935 (2000).

%121
\bibitem{Meinel_2000} I. Meinel, D. Grundler, T. Hengstmann, C. Heyn,
   D. Heitmann, W. Wegscheider, and M. Bichler, Physica E {\bf 6},
   731 (2000).

%27
\bibitem{Wang_2007a} Z. Wang and P. Zhang, Phys. Rev. B {\bf 76},
   064406 (2007).

%25
\bibitem{Wang_2007b} Z. Wang, P. Zhang, and J. Shi, Phys. Rev. B
   {\bf 76}, 094406 (2007).

%10
\bibitem{Cheng_2009} F. Cheng, W. Zhi-Gang, L. Shu-Shen, and Z. Ping,
   Chin. Phys. B {\bf 18}, 5431 (2009).

%113
\bibitem{Todorova_2001} M. Todorova, L.\,M. Sandratskii, and J.
   K\"ubler, Phys. Rev. B {\bf 63}, 052408 (2001).

%175
\bibitem{Brooks_1993} M.\,S.\,S. Brooks, O. Eriksson, L. Severin,
   and B. Johansson, Physica B {\bf 192}, 39 (1993).

%36
\bibitem{Annett_2009} J.\,F. Annett, B.\,L. Gy\"orffy, and K.\,I.
   Wysoki\'nski, New J. Phys. {\bf 11}, 055063 (2009).

%65
\bibitem{Braude_2006} V. Braude and E.\,B. Sonin, Phys. Rev. B
   {\bf 74}, 064501 (2006).

%62
\bibitem{Azimonte_2007} C. Azimonte, J.\,C. Cezar, E. Granado,
   Q. Huang, J.\,W. Lynn, J.\,C.\,P. Campoy, J. Gopalakrishnan,
   and K. Ramesha, Phys. Rev. Lett. {\bf 98}, 017204 (2007).

%100
\bibitem{Lovesey_2002} S.\,W. Lovesey, K.\,S. Knight, and D.\,S.
   Sivia, Phys. Rev. B {\bf 65}, 224402 (2002).

%159
\bibitem{Solovyev_1995} I.\,V. Solovyev, P.\,H. Dederichs, and I.
   Mertig, Phys. Rev. B {\bf 52}, 13419 (1995).

%32
\bibitem{Xiao_2006} D. Xiao, Y. Yao, Z. Fang, and Q. Niu, Phys. Rev.
   Lett. {\bf 97}, 026603 (2006).

%47
\bibitem{Autes_2008} G. Aut\`es, C. Barreteau, M.\,C. Desjonqu\`eres,
   D. Spanjaard, and M. Viret, Europhys. Lett. {\bf 83}, 17010
   (2008).

%106
\bibitem{Koide_2001} T. Koide, H. Miyauchi, J. Okamoto, T. Shidara, A.
   Fujimori, H. Fukutani, K. Amemiya, H. Takeshita, S. Yausa, T.
   Katayama, and Y. Suzuki, Phys. Rev. Lett. {\bf 87}, 257201 (2001).

%161
\bibitem{Geller_1995} M.\,R. Geller and G. Vignale, Physica B
   {\bf 212}, 283 (1995).

%163
\bibitem{Wohlman_1995} O. Entin-Wohlman, Y. Imry, A.\,G. Aronov, and
   Y. Levinson, Phys. Rev. B {\bf 51}, 11584 (1995).

%72
\bibitem{Solovyev_2005} I.\,V. Solovyev, Phys. Rev. Lett. {\bf 95},
   267205 (2005).

\bibitem{Jackson} J.\,D. Jackson, \emph{Classical Electrodynamics}
   (Wiley, New York, 1975).

\bibitem{Mauri_96a} F. Mauri and S.\,G. Louie, Phys. Rev. Lett. 
   {\bf 76}, 4246 (1996).

\bibitem{Mauri_96b} F. Mauri, B.\,G. Pfrommer, and S.\,G. Louie, 
   Phys. Rev. Lett. {\bf 77}, 5300 (1996).

\bibitem{Pickard_02} C.\,J. Pickard and F. Mauri, Phys. Rev. Lett.
   {\bf 88}, 086403 (2002).

\bibitem{Sebastiani_01} D. Sebastiani and M. Parrinello, J. Phys.
   Chem. A {\bf 105}, 1951 (2001).

\bibitem{Sebastiani_02} D. Sebastiani, G. Goward, I. Schnell, and 
   M. Parrinello, Computer Phys. Commun. {\bf 147}, 707 (2002).

\bibitem{Hirst97} L.\,L. Hirst, Rev. Mod. Phys. {\bf 69}, 607 (1997).

\bibitem{polarization_3} R. Resta, Rev. Mod. Phys. {\bf 66}, 899
   (1994).

\bibitem{Wu_01} R. Wu, \emph{First Principles Determination of
   Magnetic Anisotropy and Magnetostriction in Transition Metal
   Alloys}, Lecture Notes in Physics 580 (Springer, Berlin, 2001).

\bibitem{Sharma_07} S. Sharma, S. Pittalis, S. Kurth, S. Shallcross,
   J.\,K. Dewhurst, and E.\,K.\,U. Gross, Phys. Rev. B {\bf 76},
   100401(R) (2007).
 
\bibitem{Haldane_88} F.\,D.\,M. Haldane, Phys. Rev. Lett. {\bf 61}, 
   2015 (1988).

\bibitem{Ohgushi_00} K. Ohgushi, S. Murakami, and N. Nagaosa, 
   Phys. Rev. B {\bf 62}, R6065 (2000).

\bibitem{Jungwirth_02} T. Jungwirth, Q. Niu, and A.\,H. MacDonald,
   Phys. Rev. Lett. {\bf 88}, 207208 (2002).

\bibitem{Murakami_03} S. Murakami, N. Nagaosa, and S.-C. Zhang,
   Science {\bf 301}, 1348 (2003).

\bibitem{Yao_04} Y. Yao, L. Kleinman, A.\,H. MacDonald, J. Sinova,
   T. Jungwirth, D.-S. Wang, E. Wang, and Q. Niu,
   Phys. Rev. Lett. {\bf 92}, 037204 (2004).

%74
\bibitem{Thonhauser_05} T. Thonhauser, D. Ceresoli, D. Vanderbilt,
   and R. Resta, Phys. Rev. Lett. {\bf 95}, 137205 (2005).

\bibitem{Marzari_97} N. Marzari and D. Vanderbilt, Phys. Rev. 
   B {\bf 56}, 12847 (1997).

%28
\bibitem{Brouder_2007} C. Brouder, G. Panati, M. Calandra,
   C. Mourougane, and N. Marzari, Phys. Rev. Lett. {\bf 98},
   046402 (2007).

\bibitem{He_01} L. He and D. Vanderbilt, Phys. Rev. Lett.
   {\bf 86}, 5341 (2001).

%29
\bibitem{Thonhauser_06} T. Thonhauser and D. Vanderbilt, Phys.
   Rev. B {\bf 74}, 235111 (2006).

%76
\bibitem{Resta_05} R. Resta, D. Ceresoli, T. Thonhauser, and
   D. Vanderbilt, ChemPhysChem. {\bf 6}, 1815 (2005).

\bibitem{Gat_03b} O. Gat and J.\,E. Avron, Phys. Rev. Lett. {\bf 91},
   186801 (2003).

%31
\bibitem{Ceresoli_2006} D. Ceresoli, T. Thonhauser, D. Vanderbilt,
   and R. Resta, Phys. Rev. B. {\bf 74}, 024408 (2006).

%53
\bibitem{Souza_2008} I. Souza and D. Vanderbilt, Phys. Rev. B
   {\bf 77}, 054438 (2008).

%75
\bibitem{Xiao_2005} D. Xiao, J. Shi, and Q. Niu, Phys. Rev. Lett.
   {\bf 95}, 137204 (2005).

\bibitem{Wang_06} X. Wang, J.\,R. Yates, I. Souza, and D. Vanderbilt,
   Phys. Rev. B {\bf 74}, 195118 (2006).

\bibitem{Reichl_80} L.\,E. Reichl, \emph{A Modern Course in Statistical
   Physics} (University of Texas Press, Austin, 1980).

\bibitem{Niu_05} Q. Niu, \emph{Berry Phase and the Anomalous Hall
   Effect}, talk presented at the \emph{17$^\text{th}$ Annual Workshop
   on Recent Developments in Electronic Structure Methods} (Cornell
   University, June 24, 2005, Ithaca, New York). Available at
   \url{http://es05.physics.cornell.edu/schedule.shtml}.

%126
\bibitem{Sundaram_1999} G. Sundaram and Q. Niu, Phys. Rev. B
   {\bf 59}, 14915 (1999).

%24
\bibitem{Shi_2007} J. Shi, G. Vignale, D. Xiao, and Q. Niu, Phys. Rev.
   Lett. {\bf 99}, 197202 (2007).

\bibitem{Vignale_88} G. Vignale and M. Rasolt, Phys. Rev. B {\bf 37},
   10685 (1988).

\bibitem{Sai_02} N. Sai, K.\,M. Rabe, and D. Vanderbilt, Phys. Rev.
   B {\bf 66}, 104108 (2002).

\bibitem{Souza_04} I. Souza, J. \'{I}\~{n}iguez, and D. Vanderbilt,
   Phys. Rev. B {\bf 69}, 085106 (2004).

\bibitem{Baroni_01} S. Baroni, S. Gironcoli, A. Dal Corso, and P.
   Giannozzi, Rev. Mod. Phys. {\bf 73}, 515 (2001).

\bibitem{Bloechl_94} P.\,E. Bl\"ochl, Phys. Rev. B {\bf 50}, 17953
   (1994).

\bibitem{GIPAW} See \url{http://www.gipaw.net}.

\bibitem{Pickard_01} C.\,J. Pickard and F. Mauri, Phys. Rev. B
   {\bf 63}, 245101 (2001).

\bibitem{Schreckenbach_97} G. Schreckenbach and T. Ziegler, 
   J. Phys. Chem. A {\bf 101}, 3388 (1997).

%26
\bibitem{Ceresoli_2007} D. Ceresoli and R. Resta, Phys. Rev. B
   {\bf 76}, 012405 (2007).

\bibitem{Car_85} R. Car and M. Parrinello, Phys. Rev. Lett.
   {\bf 55}, 2471 (1985).

\bibitem{Resta_96} R. Resta, in \emph{Berry Phase in Electronic
   Wavefunctions}, Troisi\`eme Cycle Lecture Notes (Ecole Polytechnique
   F\'ed\'erale de Lausanne, Switzerland, 1996).

\bibitem{Lopez_11} M.\,G. Lopez, D. Vanderbilt, T. Thonhauser, I.
   Souza, Phys. Rev. B, under review (2011).

\bibitem{Souza01} I. Souza, N. Marzari, and D. Vanderbilt,
   Phys. Rev. B {\bf 65}, 035109 (2001).

\bibitem{Kohn_64} P. Hohenberg and W. Kohn, Phys. Rev. {\bf 136},
   B864 (1964).

\bibitem{Richard_04} R.\,M. Martin, {\it Electronic Structure, Basic
   Theory and Practical Methods} (Cambridge University Press,
   Cambridge 2004).

\bibitem{80} M. Head-Gordon, {\it Quantum Chemistry: Standard Methods
   and New Frontiers in Wavefunction Theory}, talk presented at
   the KITP conference \emph{From Basic Concepts to Real Materials}
   (University of California, November 5, 2009, Santa Barbara,
   California). Available at
   \url{http://online.kitp.ucsb.edu/online/excitcm\_c09/headgordon}.

\bibitem{PBE} J.\,P. Perdew, K. Burke, and M. Ernzerhof, Phys. Rev.
   Lett. {\bf 77}, 3865 (1996).

\bibitem{PZ81} J.\,P. Perdew and A. Zunger, Phys. Rev. B {\bf 23},
   5048 (1981).

\bibitem{Ghosh_88} S.\,K. Ghosh and A.\,K. Dhara, Phys. Rev. A
   {\bf 38}, 1149 (1988).

\bibitem{Vignale_04} G. Vignale, Phys. Rev. B {\bf 70}, 201102(R)
   (2004).

\bibitem{Runge_84} E. Runge and E.\,K.\,U. Gross, Phys. Rev. Lett.
   {\bf 52}, 997 (1984).

\bibitem{QE} P. Giannozzi, S. Baroni, N. Bonini, M. Calandra, R. Car, C.
   Cavazzoni, D. Ceresoli, G.\,L. Chiarotti, M. Cococcioni, I. Dabo, A.
   Dal Corso, S. de Gironcoli, S. Fabris, G. Fratesi, R. Gebauer, U.
   Gerstmann, C. Gougoussis, A. Kokalj, M. Lazzeri, L. Martin-Samos, N.
   Marzari, F. Mauri, R. Mazzarello, S. Paolini, A. Pasquarello, L.
   Paulatto, C. Sbraccia, S. Scandolo, G. Sclauzero, A.\,P. Seitsonen,
   A. Smogunov, P. Umari, and R.\,M. Wentzcovitch, J. Phys.: Condens.
   Matt. {\bf 21}, 395502 (2009). Available at
   \url{http://www.quantum-espresso.org}.

\bibitem{LB} D. Bonnenberg, K.\,A. Hempel, H.\,P.\,J. Wijn,
   \emph{1.2.1.2.4 Atomic Magnetic Moment, Magnetic Moment Density, g
   and g' Factor}, in \emph{1.2.1.2 Magnetic Properties}, ed.\ H.\,P.\,
   J. Wijn. Part of Landolt-B\"ornstein -- \emph{Group III Condensed
   Matter, Numerical Data and Functional Relationships in Science and
   Technology}, Vol. 19a: \emph{3d, 4d and 5d Elements, Alloys and
   Compounds}.

\bibitem{Wien2k} P. Blaha, K. Schwarz, G.\,K.\,H. Madsen, D. Kvasnicka,
   and J. Luitz, \emph{WIEN2k, An Augmented Plane Wave + Local Orbitals
   Program for Calculating Crystal Properties}. Available at
   \url{http://www.wien2k.at}.

\bibitem{Yao_2009} Y. Yao, \emph{First Principles Calculations of
   Orbital Magnetization -- Preliminary Results}, talk presented at
   the CECAM conference \emph{Orbital Magnetization in Condensed Matter}
   (EPFL, June 15, 2009, Lausanne, Switzerland). Abstract available at
   \url{http://www.cecam.org/workshop-4-303.html?presentation\_id=3575}.

\bibitem{Rabi} I.\,I. Rabi, J.\,R. Zacharias, S. Millman, and P.
   Kusch, Phys. Rev. {\bf 53}, 318 (1938).

\bibitem{NMR_encyclopedia} {\it Encyclopedia of NMR}, edited by
   D.\,M. Grant and R.\,K. Harris (Wiley, London, 1996).

\bibitem{Kutzelnigg_90} W. Kutzelnigg, U. Fleischer, and M. Schindler,
   \emph{NMR Basic Principles and Progress} (Springer, Berlin, 1990).

%5
\bibitem{Ceresoli_2010b} D. Ceresoli, N. Marzari, M.\,G. Lopez, and
   T. Thonhauser, Phys. Rev. B {\bf 81}, 184424 (2010).

%14
\bibitem{Thonhauser_2009b} T. Thonhauser, D. Ceresoli, and N. Marzari,
   Int. J. Quant. Chem. {\bf 109}, 3336 (2009).

\bibitem{VASP} G. Kresse, M. Marsman, and J. Furthm\"uller,
   \emph{Vienna Ab Initio Simulation Package, VASP the Guide}.
   Available at \url{http://cms.mpi.univie.ac.at/vasp}.

\bibitem{ABINIT} X. Gonze, B. Amadon, P.-M. Anglade, J.-M. Beuken,
   F. Bottin, P. Boulanger, F. Bruneval, D. Caliste, R. Caracas,
   M. Cote, T. Deutsch, L. Genovese, Ph. Ghosez, M. Giantomassi,
   S. Goedecker, D.\,R. Hamann, P. Hermet, F. Jollet, G. Jomard,
   S. Leroux, M. Mancini, S. Mazevet, M.\,J.\,T. Oliveira,
   G. Onida, Y. Pouillon, T. Rangel, G.-M. Rignanese, D. Sangalli,
   R. Shaltaf, M. Torrent, M.\,J. Verstraete, G. Zerah,
   and J.\,W. Zwanziger, Comp. Phys. Commun. {\bf 180}, 2582
   (2009). Available at \url{http://www.abinit.org}.

\bibitem{ADF} G. te Velde, F.\,M. Bickelhaupt, S.\,J.\,A. van
   Gisbergen, C. Fonseca Guerra, E.\,J. Baerends, J.\,G. Snijders,
   and T. Ziegler, J. Comp. Chem. {\bf 22}, 931 (2001).
   Available at \url{http://www.scm.com}.


\end{thebibliography}
\end{document}